\documentclass[twocolumn, trackchanges]{aastex631}
\usepackage{amsmath}
\usepackage{graphicx}
\usepackage{epsf}
\usepackage{color}
\usepackage{enumitem}
\usepackage{csvsimple,booktabs}
\usepackage[flushleft]{threeparttable}
\usepackage{pifont}
\usepackage{mathrsfs}
\usepackage{academicons}
\definecolor{orcidlogocol}{HTML}{A6CE39}
\begin{document}
\title{\textbf{An Investigation into the Low-Mass Fundamental Metallicity Relation in the Local and High-$z$ Universe}}
\author[0000-0003-4323-0597]{Isaac H. Laseter}
\email{Laseter@wisc.edu}
\affiliation{Department of Astronomy, University of Wisconsin-Madison, Madison, WI 53706, USA}
\author[0000-0003-0695-4414]{Michael V. Maseda}
\affiliation{Department of Astronomy, University of Wisconsin-Madison, Madison, WI 53706, USA}
\author{Andrew J.\ Bunker}
\affiliation{Department of Physics, University of Oxford, Denys Wilkinson Building, Keble Road, Oxford OX1 3RH, UK}
\author[0000-0002-0450-7306]{Alex J.\ Cameron}
\affiliation{Department of Physics, University of Oxford, Denys Wilkinson Building, Keble Road, Oxford OX1 3RH, UK}
\author[0000-0002-2678-2560]{Mirko Curti}
\affiliation{European Southern Observatory, Karl-Schwarzschild-Strasse 2, 85748 Garching, Germany}
\author[0000-0003-4770-7516]{Charlotte Simmonds}
\affiliation{Kavli Institute for Cosmology, University of Cambridge, Madingley Road, Cambridge, CB3 0HA, UK}
\affiliation{Cavendish Laboratory, University of Cambridge, 19 JJ Thomson Avenue, Cambridge, CB3 0HE, UK}


\begin{abstract}
    Recent JWST/NIRSpec observations have revealed high-$z$ star-forming galaxies depart from the Fundamental Metallicity Relation (FMR), yet the $z = 0$ FMR has not been well-characterized in the low-mass regime ($\rm log(M_{\star}/M_{\odot}) \lesssim 9$) for an appropriate comparison of low- and high-$z$ systems. We attempt to rectify this limitation through a meta-analysis, providing a local, observational comparison for future high-$z$ FMR studies. We analyzed common FMR fitting methods (minimization, parametric, non-parametric) for $\sim 700$ [OIII]$\lambda 4363$ emitters with $\rm log(M_{\star}/M_{\odot}) \lesssim 9$ at $z \sim 0$. We find no evidence of the FMR below $\rm log(M_{\star}/M_{\odot}) \lesssim 9$ through any method, suggesting that slowly-evolving, quasi-steady state gas reservoirs are not yet established. We simultaneously find a weak \textit{positive} correlation between metallicity and star formation, and that these systems are gas-rich with substantial diversity in \textit{effective yields} ($y_{\rm eff}$) spanning $\rm \sim 3~dex$. We demonstrate increasing $y_{\rm eff}$ correlates with decreasing FMR offsets, which in the context of the analytical and non-equilibrium gas models of Dalcanton et al. (2007), indicates a scenario where star formation bursts rapidly return and eject metals from the ISM prior to subsequent gas-balancing. Pristine infall diluting the ISM metal-content cannot lead to the $y_{\rm eff}$ diversity we measure, and thus is not the primary process behind FMR deviations. Our results suggest low-$\rm M_{\star}$ systems, regardless of redshift, depart from a steady-state gas reservoir shaping the canonical FMR, in which metallicity variations are primarily driven by star formation and enriched outflows. With this characterization, we demonstrate $z \gtrsim 3$ [OIII]$\lambda 4363$ systems are indeed more metal-poor than $z \sim 0$ counterparts ($\rm \Delta 12+log(O/H) = 0.3~dex$) at fixed $\rm M_{\star}$.
\end{abstract}


\defcitealias{Curti_2020}{C20}
\defcitealias{Andrews_2013}{AM13}
\defcitealias{Salim_2014}{S14}

\section{\textbf{Introduction}} \label{Introduction}

The positive scaling relation between stellar mass ($\rm M_{\star}$) and gas phase chemical abundances (\textit{metallicity}) is known as the mass-metallicity relation (MZR). \cite{Lequeux_1979} first discovered a relationship between dynamical mass and metallicity, but the difficulties in deriving $\rm M_{\star}$ necessitated subsequent studies to use $\rm M_{\star}$ proxies, such as luminosity \citep[e.g.,][]{Garnett_1987, Brodie_1991, Skillman_1989, Zaritsky_1994, Richer_1995, Garnett_1997}, which nonetheless yielded a relationship of the brightest systems being more metal-rich. As $\rm M_{\star}$ estimates improved, \cite{Tremonti_2004} established the contemporary MZR for $z = 0$ using the Sloan Digital Sky Survey (SDSS). The evolution of the MZR has been well-studied out to $z \sim 2-3$, with a clear trend of decreasing metallicity at fixed $\rm M_{\star}$ with increasing redshift \citep[e.g.,][]{Shapley_2005, Erb_2006, Maiolino_2008, Mannucci_2010, Zahid_2011, Henry_2013, Wuyts_2014, Yabe_2014, Zahid_2014, Guo_2016, Sanders_2021, Topping_2021}.

The intrinsic scatter of the MZR was shown to have strong residuals with star formation rate (SFR) \citep[e.g.,][]{Ellison_2008, Mannucci_2010, Lara_2010, Brisbin_2012, Hunt_2012, Yates_2012, Andrews_2013, Nakajima_2014, Salim_2014, Curti_2020} and gas-mass \citep[e.g.,][]{Bothwell_2013, Hughes_2013, Lopez_2013, Jimmy_2015, Brown_2018}. These additional dependencies led to the construction of the fundamental metallicity relation (FMR), which is typically presented as the inverse relationship between SFR (or gas-mass) and metallicity at fixed $\rm M_{\star}$. The FMR has been characterized through simple residual minimization methods \citep[e.g.,][]{Mannucci_2010, Andrews_2013}, principal component analysis \citep{Lara_2010, Hunt_2012, Hunt_2016}, physically-motivated parameterizations \citep{Curti_2020}, and non-parametric techniques \citep{Salim_2014}. Distinct from the MZR, the general consensus regarding FMR evolution is that it is invariant out to $z \sim 2-3$ \citep{Mannucci_2010, Cresci_2012, Yabe_2012, Cresci_2019, Sanders_2021}, in that the evolution in metallicity with redshift at a given $\rm M_{\star}$ can be explained by the concurrent evolution in SFR.

Central ideas to the MZR and FMR are quasi-steady state gas equilibrium and regulation --- gas inflow, star formation, and gas outflow continuously compensate one another, yielding a nearly constant, or slowly evolving, gas content. Considering a relation between $\rm M_{\star}$, SFR, gas-mass, and metallicity is invariant for at least the last $\rm \sim 10~Gyrs$, it is evident why quasi-steady state equilibrium is a core idea to analytical \citep[e.g.,][]{Bouche_2010, Peeples_2011, Dayal_2013, Lilly_2013, Forbes_2014, Peng_2014, Pipino_2014, Feldmamm_2015, Harwait_2015, Yabe_2015, Hunt_2016, Kacprzak_2016}, semi-analytical \citep[e.g.,][]{Somerville_2008, DeLucia_2010, Fu_2013, Hirschmann_2013, Porter_2014, Yates_2014, Cousin_2016, Zoldan_2017}, and numerical \citep[e.g.,][]{Vogelsberger_2014, Hopkins_2014, Schaye_2015, Dave_2016, Dubois_2016, Springel_2018} galaxy evolutionary models. Thus, the shape, normalization, and evolution of the MZR and FMR provides unique insights into the physical processes driving galaxy formation and evolution across cosmic time. Hence, the MZR and FMR are of high-priority for the high-$z$~community, both observationally and theoretically, considering measurements characterizing the MZR and FMR have been limited until recently to $z \lesssim 3$ due to a scarcity of near-infrared spectroscopic measurements.

The James Webb Space Telescope (JWST), particularly the Near Infrared Spectrograph \citep[NIRSpec;][]{Ferruit_2022, Jakobsen_2022, Boker_2023}, has enabled observations of the rest-frame UV/optical into cosmic dawn where the MZR/FMR can be inspected. Studies of the MZR at $z \gtrsim 3$ \citep[e.g.,][Lewis et al. in prep.]{Curti_2023, Curti_2023b, Heintz_2023, Langeroodi_2023_MZR, Matthee_2023, Nakajima_2023, Shapley_2023, Morishita_2024, He_2024, Pallottini_2025, Sarkar_2025, Scholte_2025} have generally found the normalization is offset from local MZRs ($\rm \sim 0.4-0.7~dex$), there is considerable scatter around the MZR ($\rm \sim 0.5~dex$), there is no clear evolution of the MZR above $z \sim 4$, and the slope of the MZR appears to flatten with decreasing $\rm M_{\star}$, though this flattening is dependent on the method of measuring metallicity and sample completeness. 

Interestingly, when comparing high-$z$ metallicities with locally defined FMR predictions above $z \sim 4$, individual galaxies appear to be more metal poor at fixed SFR and $\rm M_{\star}$, i.e., offset below the FMR. This offset was first identified in \cite{Heintz_2023}, and has since been corroborated in \cite{Curti_2023b}, \cite{Nakajima_2023}, \cite{Langeroodi_2023}, \cite{Pollock_2025}, \cite{Sarkar_2025}, and \cite{Scholte_2025}, all under various metallicity methodologies. The significance of the average sample offset is small ($\rm \sim0.2~dex$) and is dependent on which local FMR is chosen \citep[e.g.,][]{Andrews_2013, Curti_2020}, but statistically significant deviations are common ($\rm 20-40\%$ of sample), reflecting the $\rm \sim 1~dex$ scatter present in these measurements. In stark contrast, the residual dispersion of the local FMR is $\rm \sim 0.05~dex$, though this is likely biased low due to median binning in \cite{Mannucci_2010}, \cite{Andrews_2013} \citepalias[henceforth][]{Andrews_2013}, and \cite{Curti_2020} \citepalias[henceforth][]{Curti_2020}. Regardless, there is an approximately two order of magnitude increase in FMR residual scatter in the high-$z$ Universe, suggestive of an increase in non-equilibrium gas states in these high-$z$ systems.

However, an implicit issue in high-$z$ FMR studies is that they probe a different mass regime compared to the $z \approx 0$ FMR studies, which largely do not probe stellar masses below $\rm log(M_{\star}/M_{\odot}) \sim 8.5-9$. High-$z$ FMR inferences are thus based on a full extrapolation of the locally parameterized FMR. A key insight was demonstrated in \cite{Curti_2023b} and \cite{Scholte_2025}, in that so-called blueberries \citep{Yang_2017a} and green peas \citep{Yang_2017b}, along with low mass ($\rm log(M_{\star}/M_{\odot})\lesssim 9$) galaxies from the Dark Energy Survey Instrument \citep[DESI;][]{DESI_2024_EDR}, fall below the local FMR depending on which FMR parameterization is chosen. These results begin to offer an interesting possibility --- the high-$z$ FMR deviations are not specific to high-$z$ Universe but to low mass ($\rm log(M_{\star}/M_{\odot}) \lesssim 9$) star-forming galaxies (SFGs).

Deviations from the FMR have been tentatively observed in low-$\rm M_{\star}$ SFGs locally. \cite{Ly_2014} investigated 20 [OIII]$\lambda 4363$ emitters at $0.065 \lesssim z \lesssim0.90$ with $\rm 7 \lesssim log(M_{\star}/M_{\odot}) \lesssim 9.5$, finding $8/20$ systems fall below the \citetalias{Andrews_2013} FMR, with three systems more than $\rm 2\sigma$ away. \cite{Ly_2014} additionally finds a flat correlation between metallicity and specific star formation rate (sSFR), which is inconsistent with the FMR. \cite{Amorin_2014} and \cite{Calabro_2017} investigated star-forming dwarf galaxies ($\rm 7 \lesssim log(M_{\star}/M_{\odot}) \lesssim 9.5$), finding their samples are consistent with the \cite{Mannucci_2010} and \citetalias{Andrews_2013} FMRs but with increased dispersion greater than observational error. \cite{Calabro_2017} further demonstrated a consistent $\rm \sim 1~dex$ scatter in metallicity ($\rm 7.5 \lesssim 12+log(O/H) \lesssim 8.5$) for gas fractions ($f_{\rm gas} \rm \equiv \frac{M_{\rm gas}}{M_{\star}+M_{\rm gas}}$) ranging between $\rm \sim 0.3-0.9$, additionally suggesting larger dispersion in the low-mass FMR regime. Unfortunately, small sample sizes have limited the ability to draw a strong conclusion about the extension of the FMR to this mass range. There is a clear need to revisit the low-$\rm M_{\star}$ regime of the FMR, especially in the era of JWST and the emerging high-$z$ results. Therefore, we present a meta-analysis of the low-$\rm M_{\star}$, $z \sim 0$ FMR to provide a basis and context to high-$z$ findings. We briefly introduce novel high-$z$ FMR results later in this work, which we will expand on in a subsequent study with the \texttt{MEGATRON} simulations \citep{Katz_2024_Mtron, Katz_2025_Mtron}.

The structure of this paper is as follows: in Sect. \ref{Collected Observations}, we describe our collected sample; in Sect. \ref{Section MZR}, we give a brief overview of the MZR and FMR residuals \& extrapolations; in Sect. \ref{The FMR(s)}, we apply various FMR methodologies; in Sect. \ref{Gas-Fractions}, we investigate $f_{\rm gas}$ and \textit{effective yields} while discussing our findings; and in Sect. \ref{Summary}, we present our conclusions. For this work, we adopt the \cite{Planck_2020} cosmology: H$_0$ = 67.36 km/s/Mpc, $\rm \Omega_{m} = 0.3153$, and $\rm \Omega_{\lambda}$ $= 0.6847$.

\section{\textbf{Collected Observations}} \label{Collected Observations}

Our observational requirements for $z \approx 0$ literature in our meta-analysis are as follows:
\begin{enumerate}[nosep]
    \item $\rm[OIII]\lambda4363$ detection with $\rm S/N \geq 3$ or direct $\rm Te$-derived abundances.
    \item Measured stellar masses below $\rm log(M_{\star}/M_{\odot})\lesssim 10$.
    \item H$\alpha$ or H$\beta$ detection with $\rm S/N \geq 5$ or H$\alpha$/H$\beta$-derived SFRs.
\end{enumerate}
We therefore include the following samples: the early data release of Dark Energy Spectroscopic Instrument \citep[DESI;][]{DESI_2024_EDR}, the COS Legacy Archive Spectroscopic SurveY (CLASSY; \cite{Berg_2022}), the Spitzer Local Volume Legacy survey \citep[LVL;][]{Berg_2012}, the Hobby-Eberly Telescope Dark Energy Experiment \citep[HETDEX;][]{Indahl_2021}, the Large Sky Area Multi-Object Fiber Spectroscopic Telescope spectral survey \citep[LAMOST;][]{Gao_2017}, `Blueberries' and `Green Peas' \citep{Yang_2017a, Yang_2017b}, the Metal Abundances across Cosmic Time survey \citep[MACT;][]{Ly_2016}, and Keck data from \cite{Ly_2014}. Our total combined sample consists of $\rm 673$~galaxies.
 
An apparent shortcoming of our combined sample is the diversity of various assumptions involved in deriving $\rm M_{\star}$ (the assumed initial mass function; IMF), $\rm SFR$ (the assumed conversion of $\rm H\alpha$ to ionizing photon rate), and $\rm 12+log(O/H)$ (the atomic data) as well as dust assumptions. We do not correct to a common frame for these systematics as we find this exercise to be sub-dominant relative to the observational error and the parameter space $\rm M_{\star}$, $\rm SFR_{10}$ (the $\rm SFR$ over the past $\rm 10Myr$), and metallicity probed by this work ($\rm 6.5 \lesssim log(M_{\star}/M_{\odot}) \lesssim 10$, $\rm -2 \lesssim log(SFR_{10}) \lesssim 2$, and $\rm 7.0 \lesssim 12+log(O/H) \lesssim 8.5$). Although we do not directly address these systematic errors, we find our results remain when using self-consistent DESI results. As supporting evidence that our meta-analysis is robust, \cite{Popesso_2023} makes similar assumptions and demonstrates a minimal impact on the relation between $\rm SFR_{10}$ and $\rm M_{\star}$ relative to more self-consistent studies \citep{Speagle_2014} and Illustris-TNG300 \citep{Pillepich_2018}.

For the samples with only fluxes reported, we derive metallicities as described in \cite{Laseter_2023, Laseter_2025}. Briefly, we derive the electron temperature ($\rm T_{e}$) for O$^{++}$ by taking flux ratio of  [OIII]$\lambda\lambda4959, 5007$ doublet to [OIII]$\lambda4363$. We derive the O$^{+}$ $\rm T_{e}$ using the the O$^{++}$-O$^{+}$ $\rm T_{e}$ conversion from \cite{Curti_2017}. We used \texttt{Pyneb} \citep{Luridiana_2015} with O$^{++}$ and O$^{+}$ collision strengths from \cite{AK_1999} \& \cite{Palay_2012}  and \cite{Pradhan_2006} \& \cite{Tayal_2007}. We determined ionic oxygen abundances using \texttt{Pyneb} with the same collision strengths as before. We assume an electron density of $n_e = 300 \text{cm}^{-3}$, but the choice of electron density does not significantly affect the temperature \& metallicity results. SFRs were  derived using the conversion of \cite{Reddy_2018}, appropriate for low-metallicity galaxies. This relation has a lower $\rm L_{H\alpha}-SFR_{10}$ conversion factor than what is typically used \citep[e.g.,][see Section \ref{High-z Systems}]{Hao_2011}.

\section{\textbf{The Mass-Metallicity Relation}} \label{Section MZR}

\begin{figure*}[hbt!]
    \centering
    \includegraphics[width = \textwidth]{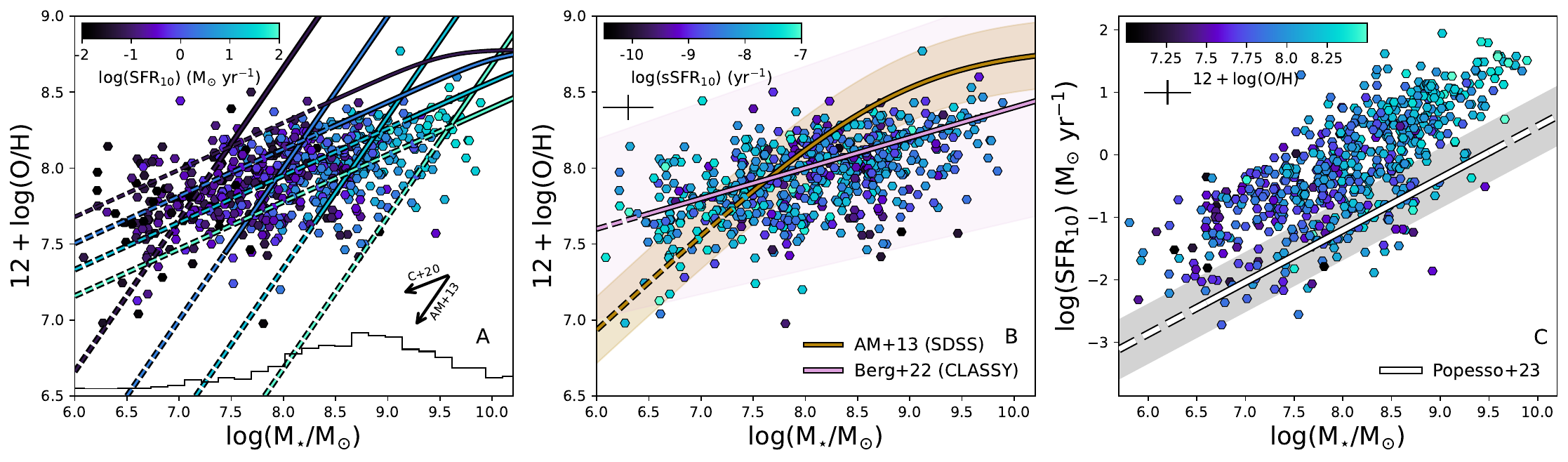}
    \caption{\textit{Panel A}: The low-$\rm M_{\star}$ MZR with $\rm SFR_{10}$ residuals. The FMRs from \citetalias{Andrews_2013} and \citetalias{Curti_2020} are presented as well, color-coded by the respective $\rm SFR_{10}$ at fixed $\rm log(M_{\star})$. Low-$\rm M_{\star}$ systems demonstrate a gradual decrease in $\rm SFR_{10}$ with $\rm log(M_{\star})$ with large O/H scatter ($\rm \sim 1dex$). We also include an inset normalized histogram of high-$z$ stellar masses from JADES, demonstrating that high-$z$ studies have been extrapolating the locally sampled FMR well over a dex to lower $\rm M_{\star}$. \textit{Panel B}: Same as panel A, but with $\rm sSFR_{10}$ residuals instead. We include the MZRs from \citetalias{Andrews_2013} and \cite{Berg_2022} (CLASSY). \textit{Panel C}: The SFMS for our low-$\rm M_{\star}$ sample, including the low-$\rm M_{\star}$ SFMS from \cite{Popesso_2023} and $\rm 12+log(O/H)$ residuals.}
    \label{MZR}
\end{figure*}

We begin with a more qualitative discussion of the low-mass FMR based on MZR residuals. We present in Figure \ref{MZR} the MZR of our combined sample, where panels A and B are color-coded by $\rm SFR_{10}$ and $\rm sSFR_{10}$, respectively. In the context of the high-$z$~Universe it is evident from the dashed lines of \citetalias{Andrews_2013} and \citetalias{Curti_2020} that high-$z$ studies have been extrapolating the locally sampled FMR well over a dex, with $\rm SFR_{10} \approx 1-10~M_{\odot}$ being equivalent to a $\rm \sim 2~dex$ extrapolation, at least when considering the JWST Advanced Deep Extragalactic Survey (JADES) reported $\rm M_{\star}$-distribution, as given by the inset normalized histogram. It is, therefore, prudent to examine the assumption these extrapolations are applicable in the low-$\rm M_{\star}$ regime, i.e., $\rm log(M_{\star}/M_{\odot}) \lesssim 9$.

It is apparent from figure panels A and B that these extrapolations are erroneous even for $z \sim 0$, at least in capturing the diversity of $\rm 12+log(O/H)$ at fixed $\rm SFR_{10}$. Although there is a clear gradient of lower SFRs at lower $\rm M_{\star}$, this is not the FMR as this trend is originating from the star-forming main sequence (SFMS) --- the FMR would be observed as low metallicity dispersion \textit{along} the SFMS, i.e., the constant SFR lines in the MZR. This low dispersion is not present in the $\rm sSFR_{10}$ residuals in panel B; at fixed $\rm sSFR_{10}$ there is $\rm \sim 1~dex$ scatter in $\rm 12+log(O/H)$, contrary to the FMR when sampled above $\rm log(M_{\star}/M_{\odot}) \approx 9$ that displays well-separated, gradual $\rm 12+log(O/H)$ changes on the order of $\rm 0.2-0.3~dex$ \citepalias{Curti_2020}. We include the SFMS with $\rm 12+log(O/H)$ residuals for our combined sample in panel C of Figure \ref{MZR}. It is clear the expected FMR trend of low metallicity dispersion along the SFMS is not present.

Although we discuss high-$z$ results in more detail in Section \ref{High-z Systems}, it is interesting that low-$\rm M_{\star}$ systems generally align with the measured high-$z$ FMR offsets and scatter reported in \cite{Curti_2023b}, \cite{Heintz_2023}, \cite{Nakajima_2023}, \cite{Langeroodi_2023}, \cite{Sarkar_2025}, and \cite{Scholte_2025} (see Section \ref{Section_Minimization}, Figure \ref{Minimization} and Section \ref{High-z Systems}, Figure \ref{high-z plot} for more detail). It is unclear, however, whether these FMR offsets are dominated by the intrinsic scatter of the population or by systematics/observational error. We explore the respective scatter-components for our $z \sim 0$ sample with a Markov Chain Monte Carlo (MCMC) algorithm using \texttt{Emcee} \citep{Foreman_2013}. We assume $\rm \sigma_{Total} = \sqrt{\sigma_{obs}^{2}+\sigma_{FMR}^{2}+\sigma_{Intrinsic}^{2}}$, where $\rm \sigma_{obs}$~represents the scatter due to $\rm 12+log(O/H)$ error, $\rm \sigma_{FMR}$~represents the scatter of the predicted FMR metallicity accounting for SFR and $\rm log(M_{\star})$ measurement error, and $\rm \sigma_{Intrinsic}$ represents the intrinsic scatter due to underlining physical processes affecting metallicity; as such, our logarithmic likelihood is defined as
\begin{equation}
    \rm ln(\mathscr{L}) = -\frac{1}{2}\sum_{i} \frac{(Z_{obs,i} - Z_{FMR,i})^{2}}{\sigma_{Total,i}^{2}}+ln(2\pi\sigma_{Total,i}^{2}).
\end{equation}
For our combined sample we find $\rm \sigma_{obs} = 0.08 \pm ^{0.08}_{0.08}$, $\rm \sigma_{FMR} = 0.14 \pm ^{0.01}_{0.01}$, and $\rm \sigma_{Intrinsic} = 0.35 \pm ^{0.01}_{0.01}$, thus the increased residual scatter is \textit{physically driven}, implying a deviation of the canonical FMR in the low-$\rm M_{\star}$ regime. Further analysis is required before any meaningful conclusions can be made, however. We now turn our attention to the various FMR methodologies. 

\section{\textbf{The FMR(s)}} \label{The FMR(s)}

The following FMR-method subsections are organized as follows: 1) minimization, 2) parametric, and 3) non-parametric. 

\subsection{\textbf{Minimization}} \label{Section_Minimization}

\begin{figure*}[hbt!]
    \centering
    \includegraphics[width = \textwidth]{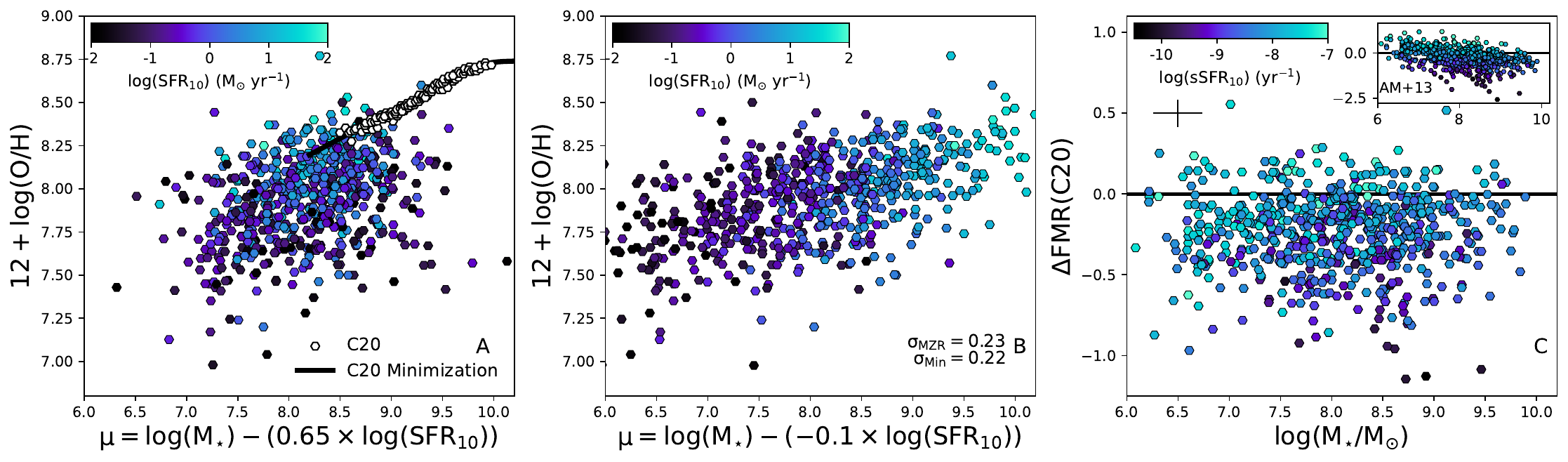}
    \caption{\textit{Panel A}: The \citetalias{Curti_2020} minimization with our low-$\rm M_{\star}$ sample overlaid with $\rm SFR_{10}$ residuals. \textit{Panel B}: Our attempt at minimizing $\rm SFR_{10}$ residuals in the low-$\rm M_{\star}$ regime. We derive $\rm \alpha = -0.1$, which is in distinct contrast to the \citetalias{Curti_2020} ($\alpha = 0.65$) and \citetalias{Andrews_2013} ($\alpha = 0.66$). Our minimization fails to decrease MZR O/H dispersion based on $\rm SFR_{10}$ residuals as $\rm \sigma_{MZR} \approx \sigma_{Min}$. \textit{Panel C}: System deviations from the \citetalias{Curti_2020} FMR per $\rm log(M_{\star})$ with $\rm sSFR_{10}$ residuals.}
    \label{Minimization}
\end{figure*}

The \cite{Mannucci_2010} and \citetalias{Andrews_2013} FMR form is simply:
\begin{equation} \label{FMR_min_scatter}
    \rm \mu_{\alpha} = log(M_{\star}) - \alpha \times log(SFR),
\end{equation}
where $\alpha$~represents the strength of secondary O/H-SFR dependency, i.e., the parameter that minimizes the MZR scatter. This parameterization is limited, however, as minimization intrinsically assumes O/H-SFR trends are $\rm M_{\star}$ invariant, which parameterized techniques disfavor.
For example, \citetalias{Curti_2020} explicitly parametrizes O/H-SFR trends (see Section \ref{Parametric}), which clearly displays a discrepancy in Figure \ref{Minimization} with \citetalias{Andrews_2013}, yet the  \citetalias{Curti_2020} minimization result ($\alpha = 0.65$) is in agreement with \citetalias{Andrews_2013} ($\alpha = 0.66$), suggesting the discrepancies in the extrapolations are due in part to the FMR technique. Regardless, we attempt to minimize our combined sample.

In panel A of Figure \ref{Minimization} we present the \citetalias{Andrews_2013} minimization with our combined sample overlaid. There is a clear discrepancy in both $\rm 12+log(O/H)$ predictions and the residual scatter, implying a deviation from $\rm SFR_{10}$ trends described by the canonical FMR. We attempt our own minimization for the low-$\rm M_{\star}$ regime, which we present in panel B of Figure \ref{Minimization}. We find an $\rm \alpha$ of $\rm -0.20$, which is in complete disagreement with prior derivations at higher $\rm M_{\star}$ \citep[e.g.,][$\alpha = 0.32, 0.19, 0.65, 0.66$, respectively]{Mannucci_2010, Yates_2012, Andrews_2013, Curti_2020}, and suggests a weak inversion of the $\rm O/H-SFR_{10}$ dependence relative to the canonical FMR. The lowest $\rm \alpha$ previously reported from \cite{Wu_2016} is $\rm 0.07$, but this was shown to primarily originate from the metallicity method employed \citep[e.g., strong-line diagnostics;][]{Sanders_2017}, meaning our deviation is substantial considering pure [O{\sc III}]$\lambda 4363$ samples correspond to higher $\rm \alpha$ values. By consequence of this technique, even after minimization, the MZR presented in Figure \ref{MZR} is approximately returned (in scatter), or rather, the minimization technique fails to reduce the $\rm 12+log(O/H)$ dispersion ($\rm \sigma_{O/H}$) based on the $\rm SFR_{10}$ dependency in the low-$\rm M_{\star}$ regime. Therefore, in the low-$\rm M_{\star}$ regime, we measure greater intrinsic $\rm \sigma_{O/H}$ (Section \ref{Section MZR}) while simultaneously being unable to reduce the underlining residuals with $\rm SFR_{10}$. In more simple framing, the minimization FMR-technique relies on an $\rm 12+log(O/H)$ dependence in the SFMS, which we demonstrated is not present in Figure \ref{MZR}.

Interestingly though, we find $\rm sSFR_{10}$ to be a moderate residual with $\rm \Delta FMR$ \citepalias{Curti_2020}, which we present in panel C of Figure \ref{Minimization}. We include $\rm \Delta FMR$ \citepalias{Andrews_2013} in the inset panel, which in the $\rm sSFR_{10}$ residuals are stronger due to the simple linear dependence between $\rm SFR_{10}$ and $\rm M_{\star}$ in the \citetalias{Andrews_2013} FMR definition. Regardless, the tendency for higher $\rm sSFR_{10}$ systems to deviate less from the canonical FMR indicates a departure of the physics relating the SFMS and the MZR (Figure \ref{MZR}), meaning gas-processes are likely deviating from equilibrium conditions that are shaping the FMR at higher $\rm M_{\star}$. There is clear dispersion in this trend, however, so additional insights from the \citetalias{Curti_2020} FMR is required before further inferences.

\subsection{\textbf{Parametric}} \label{Parametric}

\begin{figure*}[hbt!]
    \centering
    \includegraphics[width = \textwidth]{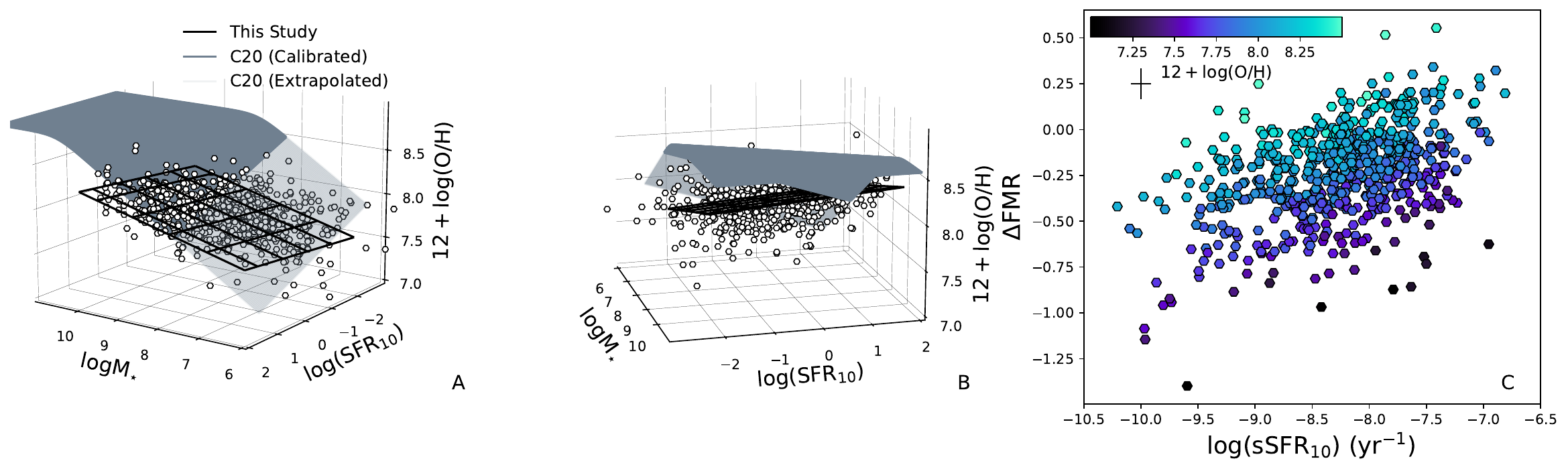}
    \caption{\textit{Panel A}: A 3D projection of the parameterized \citetalias{Curti_2020} FMR (both calibrated and extrapolated) with our low-$\rm M_{\star}$ fit and sample. The parametrized \citetalias{Curti_2020} FMR clearly intersects our sample distribution, which is relatively uniform at fixed $\rm M_{\star}$ and $\rm SFR_{10}$. \textit{Panel B}: The same distribution and relations from panel A but in a different projection. Our decreasing parameterized fit is more evident, which explains the $\rm 12+log(O/H)$ residuals within $\rm \Delta FMR$. \textit{Panel C}: $\rm sSFR_{10}$ relation with $\rm \Delta FMR$ with $\rm 12+log(O/H)$ residuals.}
    \label{3d_scatter}
\end{figure*}

The FMR proposed by \cite{Curti_2020} \citepalias{Curti_2020} explicitly accounts for O/H-SFR trends by introducing the SFR dependence within the turnover mass used in their MZR parametrization. The \citetalias{Curti_2020} MZR and FMR are given by
\begin{equation}
    \rm 12+log(O/H) = Z_{0} - \frac{\gamma}{\beta}~log(1+\frac{M}{M_{0}}^{-\beta}),
\end{equation}
\begin{equation}
    \rm 12+log(O/H) = Z_{0} - \frac{\gamma}{\beta}~log(1+\frac{M}{M_{0}(SFR)}^{-\beta}),
    \label{Curti_FMR_equation}
\end{equation}
where $\rm Z_{0}$ is the MZR/FMR metallicity saturation (i.e., the asymptotic limit for higher mass systems), $\rm M_{0}$ is characteristic turnover mass above which metallicity approaches the saturation limit, $\rm \gamma$ represents the power law for $\rm M < M_{0}$, and $\rm \beta$ represents the turnover strength, i.e., the width of the turnover at $\rm M_{0}$. The FMR is bundled in $\rm M_{0}(SFR)$, given by $\rm M_{0}(SFR)\equiv 10^{m_{0}}~SFR^{m_{1}}$, where $\rm m_{0}$ is the FMR characteristic turnover mass and $\rm m_{1}$ is the strength of the SFR dependence. We present this parametrization in panel A of Figure \ref{MZR}. It is evident that, on average, the extrapolation of \citetalias{Curti_2020}'s FMR overpredicts $\rm 12+log(O/H)$ in low-$\rm M_{\star}$ systems. \citetalias{Curti_2020} noted this systematic overestimate for a pure [OIII]$\lambda4363$ sample, though still being fully consistent within $\rm 1\sigma$. Even if we generally account for this systematic with a $\rm \sim 0.2~dex$ $\rm12+log(O/H)$ increase of our sample, this does not negate the large $\rm 12+log(O/H)$ dispersion around the constant $\rm SFR_{10}$ lines. It is therefore no longer productive to discuss residual offsets relative to \citetalias{Curti_2020} or \citetalias{Andrews_2013} since neither can capture the increased metallicity dispersion. It is also evident why the \citetalias{Andrews_2013} FMR results in extrapolations that are more in line with $\rm 12+log(O/H)$ measurements from high-z studies --- the \citetalias{Andrews_2013} results predict a steeper metallicity dependence with $\rm SFR_{10}$ at fixed $\rm M_{\star}$ than \citetalias{Curti_2020}, thus lower average metallicities are more readily predicted even when the scatter itself is not captured. 

Insights can still be gained by extrapolating the 3D \citetalias{Curti_2020} parametrization. We demonstrate in Figure \ref{3d_scatter} a 3D visualization of the \citetalias{Curti_2020} FMR (two projections with dimensions of $\rm M_{\star}-SFR_{10}-O/H$), in which it is apparent the largest FMR offsets correlate with the lowest $\rm SFR_{10}$ systems with decreasing $\rm M_{\star}$. Related, it is peculiar the high $\rm SFR_{10}$ systems posses comparable $\rm 12+log(O/H)$ values to the lower $\rm SFR_{10}$ systems and yet better align with the extrapolation of the FMR. To investigate further, we attempt to fit the low-$\rm M_{\star}$ regime with Equation \ref{Curti_FMR_equation} allowing $\rm M_{0}(SFR)$, $\rm \gamma$, and $\rm \beta$ to be free, while assuming $\rm Z_{0} = Z_{0,C20}$ as to connect with the established SDSS FMR.

We present in Figure \ref{3d_scatter} (panels A and B) our Equation \ref{Curti_FMR_equation} fit, from which it is evident $\rm \langle O/H \rangle$ is better matched, but there remains $\rm \approx 0.22~dex$ metallicity dispersion (comparable to the remaining metallicity scatter after minimization). We find our relation yields a flat O/H-$\rm SFR_{10}$ relationship at fixed $\rm M_{\star}$ that weakly becomes linear with decreasing $\rm log(M_{\star})$; these results remain if we allow $\rm Z_{0}$ to be free. We include a re-projection of this finding in panel C of Figure \ref{3d_scatter} by comparing $\rm sSFR_{10}$ with $\rm \Delta FMR$ including $\rm 12+log(O/H)$ residuals. The positive linear relation between $\rm sSFR_{10}$ and $\rm 12+log(O/H)$ is in stark contrast to the canonical FMR that predicts an inverse O/H-$\rm SFR_{10}$ relation at fixed $\rm M_{\star}$ (see Figure 6 in \citetalias{Curti_2020}). In addition, we see the strongest residual with $\rm \Delta FMR$ is $\rm 12+log(O/H)$, which represents the breakdown of the parametric FMR in the low-$\rm M_{\star}$ regime. Consequently, the highest metallicity values will better align with the lowest canonical FMR predictions, whether through a brief steady-state reservoir (the canonical FMR) or prompt metal production due to star formation in a non-equilibrium gas-state, which could simultaneously eject metals and disrupt the gas reservoir \citep{Dekel_1986, Mac_Low_1999, Woosley_2002, Dayal_2012, Hopkins_2012, Hopkins_2014, Ginolfi_2020}. It is thus likely some low-$\rm M_{\star}$ systems abide by the canonical FMR to some degree, i.e., diverse gas-balancing conditions, while some systems strongly deviate due to less metal-retention at lower $\rm SFR_{10}$, reminiscent of the $\rm sSFR_{10}$ residuals in panel C of Figure \ref{Minimization}. 

Overall, we find the standard parameters describing the high-$\rm M_{\star}$ MZR/FMR fail to accurately describe the low-$\rm M_{\star}$ FMR; we leave further discussion of why the inversion of O/H-$\rm SFR_{10}$ occurs till Section \ref{Discussion}.

\subsection{\textbf{Non-Parametric}} \label{Section Non-Parametric}

\begin{figure*}[hbt!]
    \centering
    \includegraphics[width = \textwidth]{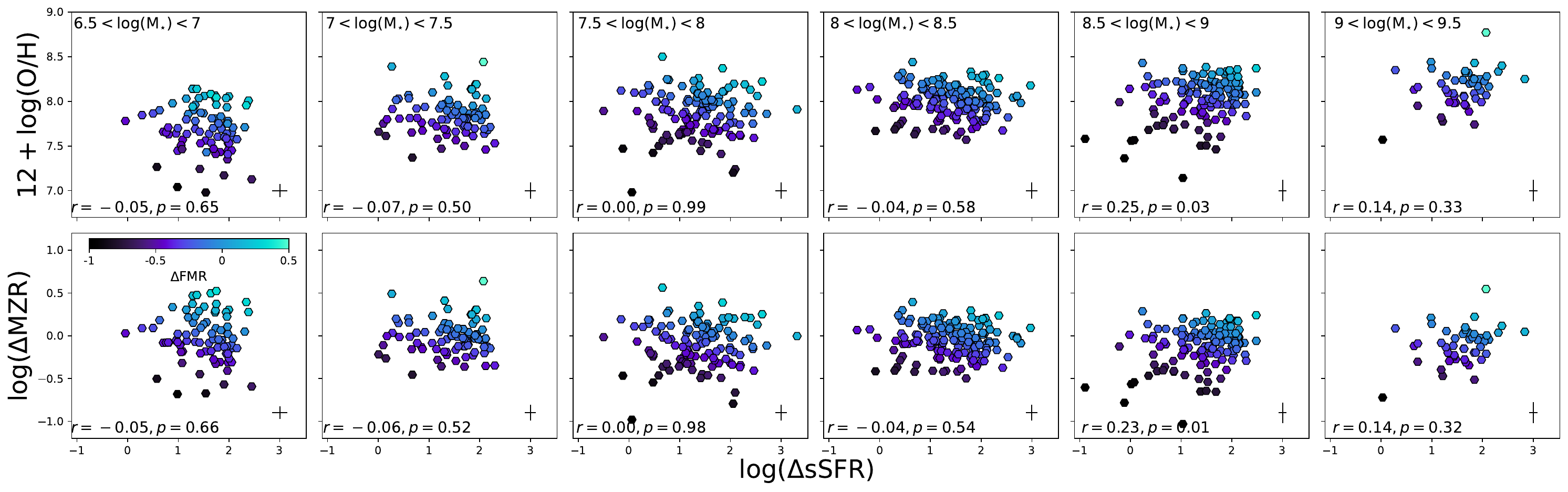}
    \caption{The non-parametric FMR as proposed by \citetalias{Salim_2014}, given by $\rm \Delta sSFR-12+log(O/H)$ (\textit{Top Row}) and $\rm \Delta sSFR-\Delta MZR$ (\textit{Bottom Row}). We parse our sample into $\rm 0.5~dex$ bins with respect to $\rm M_{\star}$, as well as including $\rm \Delta FMR$ residuals. We take $\rm \Delta sSFR$ and $\rm \Delta MZR$ relative to the low-$\rm M_{\star}$ SFMS from \cite{Popesso_2023} and the CLASSY MZR from \cite{Berg_2022}, respectively. We include Pearson r and p coefficients per bin to indicate the degree of linearity. The canonical FMR is seen as an inverse correlation that steepens with decreasing $\rm M_{\star}$, but we find uniform distributions in the low-$\rm M_{\star}$ regime, given by low r-values and high p-values. These results are simply a different parameterization of the results of Figures \ref{MZR}-\ref{3d_scatter}, further indicating a departure from a quasi-steady state gas reservoir in low-$\rm M_{\star}$ systems.   
    } 
    \label{Non-parametric}
\end{figure*}

\cite{Salim_2014} \citepalias{Salim_2014} devised a non-parametric FMR framework based on the offset from SFMS ($\rm \Delta sSFR$) relative to $\rm 12+log(O/H)$. \citetalias{Curti_2020} also investigated this avenue, but with the correlation of $\rm \Delta sSFR$ with the relative offset of the MZR ($\rm \Delta MZR$). Regardless, in this context, the FMR is seen as a flat relationship at lower SFMS offsets, with a steepening dependence the further above the SFMS and lower mass one probes. \cite{Salim_2014} investigated mass bins down to $\rm log(M_{\star}/M_{\odot}) \approx 9$, while \citetalias{Curti_2020} chose a finer binning scheme down to $\rm log(M_{\star}/M_{\odot}) \approx 8.25$. However, samples were not explored past $\rm \approx 1.5$$\rm \Delta sSFR$, and thus the regime of low-$\rm M_{\star}$ SFGs has not been well-characterized under this FMR context. If the canonical FMR is maintained at these lower masses and SFRs, we would expect the steepening found by \cite{Salim_2014} and \citetalias{Curti_2020} to persist with both decreasing $\rm M_{\star}$ and increasing $\rm SFR_{10}$. However, from our analyses in Section \ref{Section_Minimization} and \ref{Parametric}, we demonstrated a degradation in the physical processes relating the MZR and SFMS, and thus we do not necessarily expect a strong non-parametric relationship in the low-$\rm M_{\star}$ regime.  

We present in Figure \ref{Non-parametric} the $\rm \Delta sSFR$-$\rm \Delta MZR$ and $\rm \Delta sSFR$-$\rm 12+log(O/H)$ for our combined sample in mass bins of $\rm 0.5~dex$ (our results remain unchanged when employing a finer binning scheme similar to \citetalias{Curti_2020}). We also include the respective $\rm \sigma_{O/H}$ and Pearson r\&p-values (i.e., the degree of linearity) in each bin. $\rm \Delta sSFR$ is taken with respect to the SFMS from \cite{Popesso_2023}  and $\rm \Delta MZR$ is taken with respect to the CLASSY-MZR from \cite{Berg_2022}; interchanging these relations with appropriate alternatives \citep[e.g.,][]{Speagle_2014, Berg_2012} has no effect on our results. 

We find no correlation between the location of a low-$\rm M_{\star}$ SFG on the SFMS to its location on the MZR, unlike the inverse correlation identified in \cite{Salim_2014} and \citetalias{Curti_2020}. In fact, although likely due to sample variance within our binning scheme, the $\rm 8.5 \lesssim log(M_{\star}) \lesssim 9.0$ bin suggests a statistically significant \textit{linear} correlation, reminiscent of our parametric results from the preceding subsection. We find $\rm \sigma_{O/H}$ is consistent across our mass bins, with Pearson values suggesting no statistically significant linear anti-correlation. Our findings are more closely aligned to the highest-mass bins ($\rm log(M_{\star}/M_{\odot}) \gtrsim 10.5$) of \citetalias{Salim_2014} and \citetalias{Curti_2020} where the $\rm \Delta sSFR$-$\rm \Delta MZR(12+log(O/H))$ relations are featureless due to these systems reaching $\rm Z_{0}$. As noted by \citetalias{Curti_2020} for the high-$\rm M_{\star}$ regime, $\rm Z_{0}$ is regulated by the \textit{effective yield} (see Section \ref{Section Effective Yields}) and is independent of $\rm SFR_{10}$ due to the larger potential wells, yielding an asymptotic, uniform distribution. However, this explanation is not applicable in the low-$\rm M_{\star}$ regime considering there is no $\rm Z_{0}$ associated with the weaker potential wells and there is a clear $\rm \Delta FMR-sSFR_{10}$ relationship in Figure \ref{3d_scatter}, which manifests as strong $\rm \Delta FMR$ residuals in $\rm \Delta sSFR$-$\rm \Delta MZR~(12+log(O/H)$ space. Instead, these residuals suggest the \textit{relative level} of chemical enrichment is more closely related to the FMR offset, thus diminishing the importance of the \textit{relative position} above the SFMS and increasing the respective $\rm 12+log(O/H)$ dispersion. 

Overall, our non-parametric findings corroborate the minimization and parametric results, suggesting a scenario where star formation rapidly returns metals to the ISM, followed by feedback expelling metals from the ISM prior to subsequent gas-balancing, and thus a departure from the canonical FMR in the low-$\rm M_{\star}$ regime. As these low-$\rm M_{\star}$ systems are likely gas-rich \citep[][see Section \ref{Section Gas-Fractions}]{Henkel_2022}, it is clear we need to investigate gas fractions and the relative level of enrichment to the gas reservoir, i.e., effective yields (Section \ref{Section Effective Yields}), for our sample. We now turn our attention to these parameters as well as expanding our discussion.

\section{\textbf{Discussion: Gas-Fractions, Effective Yields, and High-z Observations}} \label{Discussion}

\subsection{\textbf{Gas-Fractions}} \label{Section Gas-Fractions}

\begin{figure*}[hbt!]
    \centering
    \includegraphics[width = \textwidth]{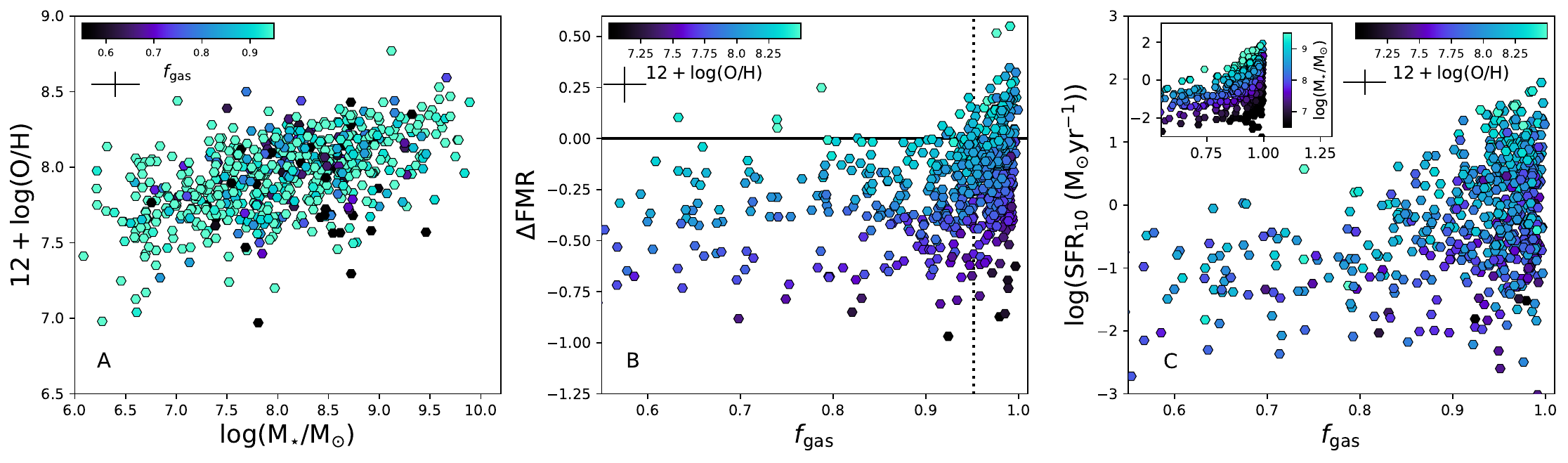}
    \caption{\textit{Panel A}: The MZR with $f_{\rm gas}$ residuals. Similar to the MZR with $\rm SFR_{10}$ and $\rm sSFR_{10}$ residuals, there is no correlation between $f_{\rm gas}$ at fixed $\rm M_{\star}$ with $\rm 12+log(O/H)$, which is contrast to the canonical FMR. \textit{Panel B}: $\rm \Delta FMR$ correlation with $f_{\rm gas}$ with $\rm 12+log(O/H)$ residuals. The majority of our low-$\rm M_{\star}$ sample is extremely gas rich, as expected; we include the median $f_{\rm gas}$ ($\rm 0.95$) as a vertical line. We find no correlation between $f_{\rm gas}$ and $\rm \Delta FMR$, though the strong relationship between $\rm \Delta FMR$ and $\rm 12+log(O/H)$ is present. \textit{Panel C}: $\rm SFR_{10}$ correlation with $f_{\rm gas}$ with $\rm 12+log(O/H)$ residuals.}
    \label{Gas-Fractions}
\end{figure*}

So far, our analysis has been focused on the FMR in the context of $\rm M_{\star}-SFR_{10}-O/H$, and thus we have neglected the role of $f_{\rm gas}$. All gas equilibrium models predict a O/H-$f_{\rm gas}$ dependence due to the increase of $\rm SFR_{10}$ with gas content and possible metal dilution, with several models \citep[e.g.,][]{Lagos_2016, Segers_2016, De_Rossi_2017} suggesting $\rm M_{\star}-\it{f_{\rm gas}} \rm -O/H$ to be more fundamental \citep{Maiolino_2019}.  \cite{Peeples_2008, Peeples_2009} showed that such a correlation is present in the local Universe, in the sense that SDSS galaxies with lower $f_{\rm gas}$ also have metallicities above the MZR. In the formalism of the FMR, \cite{Bothwell_2013}, \cite{Hughes_2013}, \cite{Lara_2013}, \cite{Jimmy_2015}, and  \cite{Brown_2018} showed that metallicities anti-correlate with $f_{\rm gas}$, and that this correlation is tighter relative to $\rm SFR_{10}$. Interestingly, \cite{Bothwell_2013} demonstrated the O/H-$f_{\rm gas}$ relation continues into higher $\rm M_{\star}$, where metallicity does not depend on SFR anymore. It is therefore essential to investigate $\rm M_{\star}-\it{f_{\rm gas}} \rm -O/H$ in our meta-analysis.

There are several limitations and caveats in obtaining $f_{\rm gas}$, however. We first attempt to use HI measurements from the Arecibo Legacy Fast ALFA Survey (ALFALFA), a blind HI survey of the local Universe \citep{Giovanelli_2005, Haynes_2018} with optical counterpart matching from \cite{Haynes_2018}. We match our sample by position to \cite{Haynes_2018} within a $\rm 5''$ threshold, retrieving $f_{\rm gas}$ measurements for $\rm 22/792~(\sim 3\%)$ of our sample, which is expected but nonetheless a substantial loss. However, the median positional offset between the optical counterpart and the HI centroid in \cite{Haynes_2018} is $\sim 18''$, hence we are probing HI masses not necessarily coincident with star formation. A possible solution is to simply assume a fixed fraction of the gas mass to be associated with star formation \citep{Li_2025_gas}.

However, in Section \ref{Section Effective Yields} we introduce the \textit{effective yield} in the context of gas-regulation models, meaning there is an implicit assumption of total gas inflow and outflow effects that is not necessarily captured in sole $\rm HI$ measurements. Unfortunately, we could not identify simultaneous observations of $\rm H_{mol}$ for our $\rm HI$ matched sample.
Turning to empirical relations, \cite{Hagedorn_2024} provides $\rm M_{\star}-\it f_{\rm mol}$ and $\rm M_{\star}-\it f_{\rm HI}$ relations, though only probing down to $\rm log(M_{\star}/M_{\odot}) \approx 8.5$, meaning it is uncertain whether these relations hold at the lower-$\rm M_{\star}$ probed in our analysis, combined with the implicit $\rm \alpha_{CO}$ conversions that are metallicity dependent. \cite{Boselli_2014} also provides combined $\rm M_{\rm gas}$ conversions, including the common factor accounting for Helium, in relation with $\rm M_{\star},~ log(sSFR_{10})$, and $\rm 12+log(O/H)$. Similar to \cite{Hagedorn_2024} though, the applicability of these relations remains uncertain in the low-$\rm M_{\star}$ regime. 

We continue with deriving $f_{\rm gas}$ from the extrapolation of the \cite{Boselli_2014} relation with respect to $\rm sSFR_{10}$, but our inferences do not vary when interchanging the other parameters or the \cite{Hagedorn_2024} relations since all ultimately yield high gas-fractions ($f_{\rm gas} \gtrsim 0.8$). We also find these empirical relations are generally consistent with the ALFALFA-matched subsample when assuming anywhere from $\rm 10-60\%$ of the HI-content being coincident with star formation. Ultimately, whether we decide on 1) sole $\rm HI$ measurements from the extrapolations of \cite{Hagedorn_2024} or 2) the extrapolations of total gas-mass from \cite{Boselli_2014}, our inferences in the following sections are based on effective yield trends, which we find are consistent regardless of the various options presented here.

\cite{Geha_2006}, \cite{Saintonge_2007}, \cite{Karachentsev_2019}, and \cite{Scholte_2024} demonstrated a $\rm \approx 1.5~dex$ scatter with $\rm log(\it{f_{\rm gas}})$ below $\rm log(M_{\star}/M_{\odot}) \lesssim 8-9$. In contrast, similar to flux-limited samples, we find higher average gas-fractions ranging between ($\rm \approx 0.6-0.9$) for our sample, with a median value of $\rm 0.95$. This observational bias is expected considering our pure [OIII]$\lambda 4363$ emitter sample, but the notion that higher $f_{\rm gas}$ correlates with lower $\rm 12+log(O/H)$ should remain if the canonical FMR is well described in the low-$\rm M_{\star}$ regime. We present in panel A of Figure \ref{Gas-Fractions} the MZR with $f_{\rm gas}$ residuals. We find no clear residuals or inverse correlation between $\rm 12+log(O/H)$ and $f_{\rm gas}$ (and likewise for $f_{\rm gas}-\rm \Delta MZR$), supported by the lack of linearity for O/H-$f_{\rm gas}$ (Pearson: $\rm r = -0.029, p = 0.451$). We also present in panel B of Figure \ref{Gas-Fractions} $\rm \Delta FMR - \it f_{\rm gas}$, where it is evident there is no correlation. We color-code panel B by $\rm 12+log(O/H)$ to further demonstrate the strong residuals $\rm \Delta FMR$ has with the level of chemical enrichment, unlike $\rm SFR_{10}$ and $f_{\rm gas}$, i.e., the canonical FMR.

We demonstrate $\rm SFR_{10}-\it f_{\rm gas}$ with $\rm 12+log(O/H)$ residuals in panel C, which in it is apparent that our systems are gas-rich regardless of the level of star formation. This is expected for low-$\rm M_{\star}$ systems \citep[e.g.,][]{Geha_2006, De_Rossi_2013,Bradford_2015, Somerville_2015, Somerville_2015_review}, but it is nonetheless peculiar that when we color-code panel C by $\rm M_{\star}$ (inset panel), the SFMS is clearly returned with the expected trends of $f_{\rm gas}$. These discrepancies align with the parametric and non-parametric results, but with the addition that $f_{\rm gas}$, obviously a dependent variable in the SFMS, does not align with the canonical FMR. These results suggest metallicity variations are on a shorter timescale than the replenishing/balancing of the gas reservoir, i.e., metallicity variations without a near-constant gas-inflow.

For instance, in low-$\rm M_{\star}$ starbursts, prompt metal injection will begin through stellar winds followed by type-II supernovae explosions \citep[SNe;][]{Chevalier_1985, Marlowe_1995, Hopkins_2011, Bolatto_2013}, and combined with the scenario there is no quasi-steady state gas content, whether through unbounding the gas \citep[e.g.,][]{Bolatto_2013, Chisholm_2017, Walter_2017, Romano_2023} or gas heating \citep[e.g.,][]{Madden_2000, Melioli_2004, Ostriker_2010, Cormier_2012, Forbes_2016}, $\rm 12+log(O/H)$ will be measured close to nucleosynthetic values. This susceptibility is even captured in the equilibrium framework from \cite{Lilly_2013} as their metallicity solution for a steady-state reservoir approaches the true nucleosynthetic yield when inflow is disrupted. Following this enrichment, outflows will begin to remove enriched gas and metallicity will decrease by an amount dependent on the intensity of the starburst \citep{Dave_2011}, proportion of metals returned, surrounding gas column densities, the gravitational well, and the IMF, leading to an increase in the intrinsic metallicity dispersion and less metal retention than predicted by the canonical FMR at lower $\rm sSFR_{10}$. In view of these physical processes, real low-$\rm M_{\star}$ systems span an array of star formation and outflow conditions, the resulting outcome is the highest metallicity values will align closest to the lowest FMR-predicted metallicity, hence the strong $\rm 12+log(O/H)$ residuals with $\rm \Delta FMR$. This physical picture also predicts a metallicity detachment in well-described gas processes still present at these masses, which we observe in the SFMS (Figure \ref{MZR}).

Implicit within low-$\rm M_{\star}$ starbursts though is the accretion of pristine gas, which may increase the gas content while diluting the metal content \citep[e.g.,][]{Dekel_2009, Lilly_2013, Somerville_2015, Somerville_2015_review, Dave_2017}. Considering our low-$\rm M_{\star}$ systems are gas rich regardless of the degree of star formation (panel C, Figure \ref{Gas-Fractions}), large metallicity variations are not expected from dilution. However, our analysis thus far has solely examined metallicity variations in the context of $\rm SFR_{10}$, $\rm M_{\star}$ and $f_{\rm gas}$, which does not suffice in fully substantiating the picture that star formation and subsequent outflows are the primary drivers of the metallicity variations in a non-steady state gas reservoir. We therefore turn our attention to examining metallicity \textit{relative} to the overall gas content, i.e., \textit{effective yields} in the context of the \cite{Dalcanton_2007} models. 

\subsection{\textbf{Effective Yields}} \label{Section Effective Yields}

\begin{figure*}[hbt!]
    \centering
    \includegraphics[width = \textwidth]{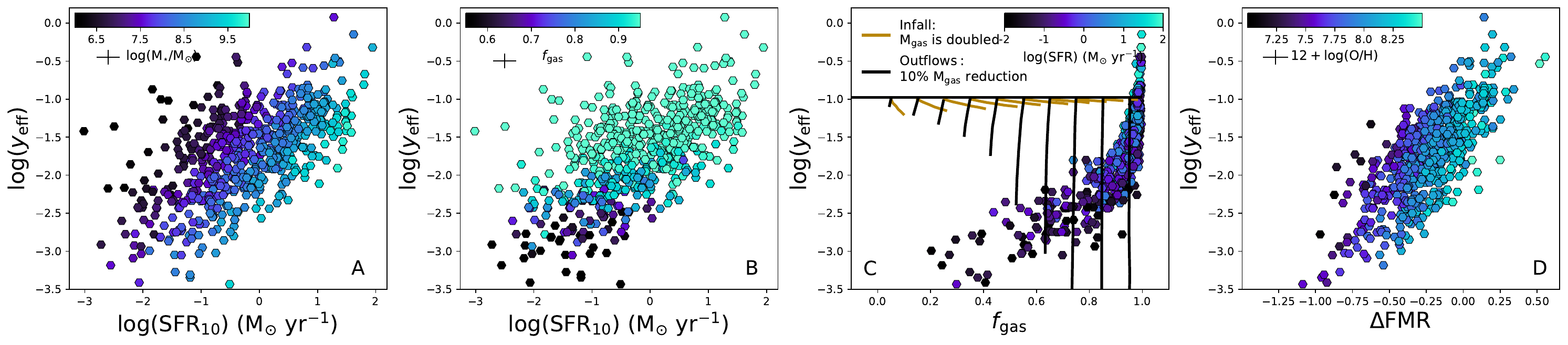}
    \caption{\textit{Panels A \& B}: $y_{\rm eff}-\rm SFR_{10}$ with $\rm M_{\star}$ and $f_{\rm gas}$ residuals, respectively. The residuals of these panels demonstrate the expected trends of the SFMS, meaning the linear $y_{\rm eff}-\rm SFR_{10}$ relation reflects the increase of metallicity with $\rm SFR_{10}$ in gas-rich systems: \textit{Panel C}: $\it y_{\rm eff} - f_{\rm gas}$ with $\rm SFR_{10}$ residuals. We include \cite{Dalcanton_2007} models for pristine infall (gold) and enriched outflows models (black) at a given $\rm log(\it y_{\rm eff}\rm) = -1$ and various $f_{\rm gas}$ values. The infall model is for the extreme case of doubling the gas reservoir, while the enriched outflow model is only assuming $\rm 10\%$ of the metal-loaded gas is removed. \textit{Panel D}: $y_{\rm eff}-\rm \Delta FMR$ with $\rm 12+log(O/H)$ residuals. This relation further indicates the increase of metallicity with $\rm SFR_{10}$ in gas-rich systems considering the strongest residual with $\rm \Delta FMR$ is $\rm 12+log(O/H)$.}
    \label{Effective Yields}
\end{figure*}

Galaxies that individually evolve as a ``closed-box'' follow a relationship between metal- and gas content given by \begin{equation}
    \rm Z_{gas, M} = \it y_{\rm true}~\rm ln(1/f_{\rm gas}),
\end{equation}
where $\rm Z_{gas, M}$ is the metallicity (abundance by mass not number) and $y_{\rm true}$ is the true nucleosynthetic yield (the elemental-yield mass produced by young, massive stars, in units of the remaining $\rm M_{\star}$, and in this case referring to Oxygen). The \textit{effective yield} is similarly defined as 
\begin{equation}
    y_{\rm eff} = \rm \frac{Z_{gas, M}}{ln(1/\it f_{\rm gas})},
    \label{Effective Yield Equation}
\end{equation}
and thus the comparison of $y_{\rm true}$ and $y_{\rm eff}$ measures the offset between the true nucleosynthetic yield and the measured metallicity due to gas and star formation processes, e.g., inflows and outflows, with $y_{\rm true} = y_{\rm eff}$ being representative of a closed-box model. For our purposes, it is not necessary to assume $y_{\rm true}$, but we note this parameter has strong dependence on the assumed IMF \citep{Vincenzo_2016_IMF}, massive star treatment \citep{Maeder_1981}, stellar metallicity \citep{Portinari_1998}, and mass-loss assumptions \citep{Chiosi_1986}. We care only for the relative trends of $y_{\rm eff}$, though $y_{\rm eff}$ does not immediately reveal why a system is in a $y_{\rm true} \neq y_{\rm eff}$ state. 

We therefore combine our analysis with the analytic calculations from \cite{Dalcanton_2007} focusing on the distinguishing factors between infall, star formation, and outflow with $y_{\rm eff}$, $y_{\rm true}$, $f_{\rm gas}$, and $\rm 12+log(O/H)$. \cite{Dalcanton_2007} did not assume a balance between gas-processes, meaning the $y_{\rm eff}$ and $\rm 12+log(O/H)$ changes are for the dominance of inflows, star formation, and outflows in the scenario there is no steady-state gas reservoir. We briefly summarize the findings of \cite{Dalcanton_2007}: In high $f_{\rm gas}$ systems, metal-enriched outflows are \textit{effective} in significantly reducing $y_{\rm eff}$ without changing the overall $f_{\rm gas}$, thus allowing large metallicity changes without changing the total gas content. In contrast, unenriched inflows and outflows are \textit{ineffective} in lowering $y_{\rm eff}$ in gas-rich systems. In the opposite regime of low $f_{\rm gas}$ systems, enriched outflows are \textit{ineffective} at reducing the $y_{\rm eff}$, and there is a minimum $y_{\rm eff}$ that can be produced by gas accretion alone. Crucially, any level of star formation will drive $y_{\rm eff} \rightarrow y_{\rm true}$, and thus galaxies with the lowest $y_{\rm eff}$ must also have elevated $f_{\rm gas}$ and low SFRs, but will rapidly approach $y_{\rm true}$ in the presence of star formation.

We present in Figure \ref{Effective Yields} $y_{\rm eff}$ against $\rm SFR_{10}$, $f_{\rm gas}$, and $\rm \Delta FMR$ with $\rm M_{\star},~\it f_{\rm gas},$ $\rm SFR_{10}$, and $\rm 12+log(O/H)$ residuals. It is evident in panels A and B that $y_{\rm eff}$ scales with $\rm SFR_{10}$, though we have introduced a clear systematic considering we determined $f_{\rm gas}$ with $\rm sSFR_{10}$. Even so, the strong $\rm M_{\star}$ and $f_{\rm gas}$ residuals with $\rm SFR_{10}$ demonstrate the expected trends of the SFMS, in that more gas-rich systems correspond to higher $\rm SFR_{10}$ at fixed $\rm M_{\star}$. As such, the predicted $y_{\rm eff}-\rm SFR_{10}$ scaling from \cite{Dalcanton_2007} is present in our sample outside the systematics we introduce. The \cite{Dalcanton_2007} models predict that in gas-rich systems the higher metallicities are primarily driven by more recent star formation, thus increasing $y_{\rm eff}$ to $y_{\rm true}$. This trend is apparent in panel D as $y_{\rm eff}$ scales with $\rm \Delta FMR$, aligning with the $\rm O/H-\Delta FMR$ residuals identified throughout this work. Likewise, our lowest $y_{\rm eff}$ values are correlated with the largest offsets of the FMR, which is predicted by \cite{Dalcanton_2007} to be primarily associated with enriched outflows than with pristine gas-accretion for gas-rich systems, which is apparent in the highly contrasting pristine infall and enriched outflow models given in Panel C. 

This result indicates low-$\rm M_{\star}$ systems are experiencing a decrease in their $\rm SFR_{10}$ coincident with enriched outflows that drives $y_{\rm true} \rightarrow y_{\rm eff}$.
Therefore, these $y_{\rm eff}-\rm \Delta FMR$ trends in the context of the \cite{Dalcanton_2007} models, combined with our failure to capture the canonical FMR through any standard approach (Section \ref{The FMR(s)}), suggests the following picture: beginning below $\rm log(M_{\star}/M_{\odot}) \lesssim 9$, there is a general departure from a steady-state gas reservoir shaping a $\rm M_{\star}-SFR_{10}-O/H$ and $\rm M_{\star}-\it f_{\rm gas}-\rm O/H$ relation, in which metallicity variations and FMR deviations are primarily driven by star formation and enriched outflows. 

\subsubsection{\textbf{Supplemental Support (Low-$z$)}}

\cite{Li_2025} found systems with irregular gas velocity fields are more likely to posses \textit{positive} metallicity gradients than those with regular velocity fields (i.e., an increase in $\rm 12+log(O/H)$ with galactocentric distance), meaning the ISM is heavily disturbed with significant metal redistribution. Our systems likely have irregular velocity distributions as \cite{Lelli_2014} found the HI morphologies of nearby low-$\rm M_{\star}$ SFGs are heavily disturbed with major asymmetries and offsets from the stellar component. Likewise, the FIRE simulation \citep[e.g.,][]{Ma_2017a, Ma_2017b} predicts irregular gas velocities associated with bursty star formation in $\rm log(M_{\star}/M_{\odot}) \lesssim 8$ systems, with $\rm 8 \lesssim log(M_{\star}/M_{\odot}) \lesssim 10$ systems displaying a wide range of gas kinematics and morphologies, i.e., the mass regime for the balancing of the feedback-driven turbulent ISM pressure and gas pressure. Although we do not directly address gas velocities for our systems, \cite{Li_2025} also demonstrated the highest/lowest $y_{\rm eff}$ values correspond to more negative/positive metallicity gradients, which conforms with our $y_{\rm eff}-\rm SFR_{10}$ and $y_{\rm eff}-\rm \Delta FMR$ relations, in that as star formation drives $y_{\rm eff} \rightarrow y_{\rm true}$ in the nuclear regions of a galaxy, the metal-spatial distribution will produce a negative gradient. In contrast, as $y_{\rm eff}$ is lowered primarily through outflows, gradients flatten or invert as the enriched gas moves to the system outskirts and $y_{\rm true} \rightarrow y_{\rm eff}$ arises. Finally, observational evidence is given by \cite{Chisholm_2018} and \cite{Hamel_Bravo_2024}, as both studies found the metal-loading factor inversely scales with $\rm M_{\star}$ with $3\sigma$ significance, meaning enriched outflows in low-$\rm M_{\star}$ galaxies are substantially more efficient at removing metals. \cite{Chisholm_2018} also demonstrated entrainment factors of low-$\rm M_{\star}$ systems disproportionally contain SNe ejecta. These types of observations are necessary to expand upon for future studies.

Considering the strongest residual with the $\rm \Delta FMR$ is $\rm 12+log(O/H)$, it is clear from the residuals of panels C of Figure \ref{Effective Yields} that some systems have higher metallicities associated with lower $y_{\rm eff}$. These systems still align with our non-equilibrium gas picture as real low-$\rm M_{\star}$ systems will encompass diverse and interrelated gas-reservoir balancing conditions, outflow properties, and recycling characteristics, and thus $\rm 12+log(O/H)$ may remain elevated and diverse, forcing $\rm \Delta FMR \rightarrow 0$. For example, variations in SNe energy injection conditions \citep[e.g., IMF and SNe clustering;][]{Smith_2021} with different mass loads dependent on ISM entrainment factors and the ISM phases encountered \citep{Chisholm_2018_outflows}, may result in varied $\rm 12+log(O/H)$ measurements on differing timescales post starburst. Additionally, the outflows themselves are multiphase phenomena \citep{Strickland_2000}. The hot phase ($\rm \gtrsim 10^{7}~K$) has been shown to be radiative if substantially mass-loaded \citep{Chisholm_2018_outflows}, thus reducing the mass outflow rate and increasing metallicity diversity. Additionally, some low-$\rm M_{\star}$ systems may simply be close to equilibrium in their respective conditions, thus leading to minor differences from the canonical FMR \citep{Forbes_2014}.

Regardless, the combination of the failure of the standard FMR analyses, the simultaneous measurements of extremely low and high $y_{\rm eff}$ values aligning with the \cite{Dalcanton_2007} non-steady-state gas reservoir models, and the suite of external observation and analytical findings of low-$\rm M_{\star}$ SFGs, supports our notion that the FMR is not fully applicable in the low-$\rm M_{\star}$ regime. Consequently, as we proposed in Section \ref{Introduction}, the significant FMR offsets measured in the high-$z$ Universe are not fully specific to the high-$z$ Universe but rather the low-$\rm M_{\star}$ regime probed. We are now in the position to directly compare the high-$z$ results within the same mass regime at $z \sim 0$.   

\subsection{\textbf{High-z Systems}} \label{High-z Systems}

\begin{figure*}[hbt!]
    \centering
    \includegraphics[width = \textwidth]{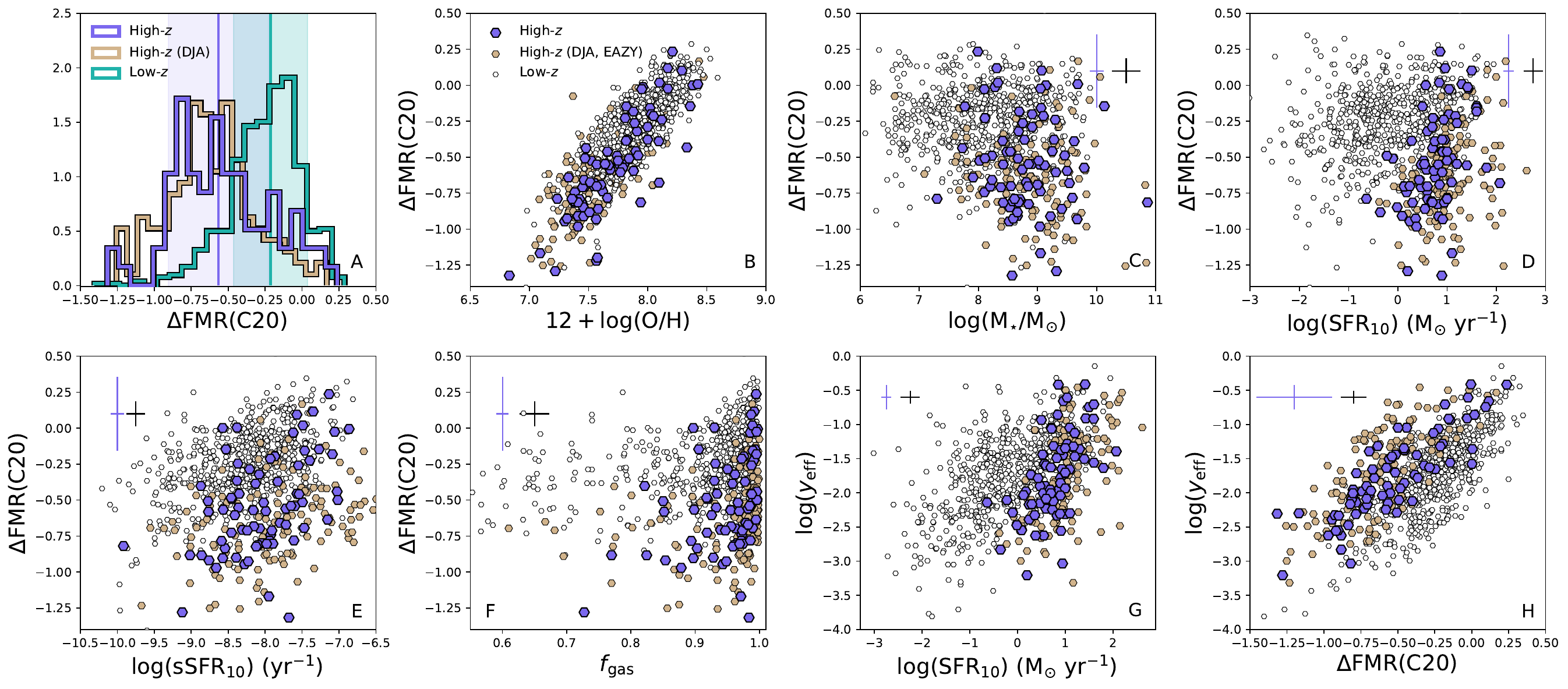}
    \caption{Comparison between low- and high-$z$ systems in parameter spaces explored thus far. It is apparent in \textit{Panels A, D}, and  \textit{E} that high-$z$ systems are disproportionally more metal-poor ($\rm \langle 12+log(O/H) \rangle = 7.66; \langle \Delta FMR \rangle = -0.57$) than low-$z$ counterparts ($\rm \langle 12+log(O/H) \rangle = 7.96; \langle \Delta FMR \rangle = -0.22$). Additionally, in \textit{Panels G} and \textit{H}, $y_{\rm eff}$ values are similar between low- and high-$z$ systems, suggesting star formation and enriched outflows still dominate the metallicity variations in the high-$z$ Universe when non-equilibrium, gas-rich reservoirs are present.}
    \label{high-z plot}
\end{figure*}

We noted in Section \ref{Introduction} the peculiarity that the severity of high-$z$ FMR deviations are contingent on which local FMR is chosen, even though the overall scatter remains comparable. We now see this level of discussion of the high-$z$ FMR is limited as any low-$\rm M_{\star}$ system with $\rm \Delta FMR \approx 0$ is serendipitous since neither of these parameterizations hold in the local Universe for the same $\rm M_{\star}$ regime, hence why the degree of scatter was invariant between the methods. However, with the low-$\rm M_{\star}$ regime better characterized in FMR frameworks, we are in a position to directly compare high-$z$ systems to low-$z$ counterparts.  

We therefore gather all $z \geq 2$ [OIII]$\lambda 4363$~emitters from the JWST Advanced Deep Extragalactic Survey \citep[JADES;][]{Bunker_2023, Eisenstein_2023, DEugenio_2024}, as well as any literature [OIII]$\lambda 4363$~emitter with a derived $\rm M_{\star}$ and $\rm SFR_{10}$ that is not directly identified with an AGN component \citep[i.e.,][]{Nakajima_2023, Sanders_2023b, Morishita_2024, Cullen_2025, Scholte_2025}. We further include all $z \geq 2$ systems from the DAWN JWST Archive \citep[DJA\footnote{https://dawn-cph.github.io/dja/}\footnote{https://zenodo.org/records/8319596};][]{Heintz_2024_DJA} with strong [OIII]$\lambda4363$ and $\rm H\beta$ emission ($\rm S/N \geq 5$), but we separate this sample as no AGN identification was done and $\rm M_{\star}$ was derived through photometry with \texttt{EAZY} alone. General SED assumptions of our JADES sample are given in \cite{Simmonds_2024, Simmonds_2025_SED}, but briefly, \texttt{Prospector} \citep{Johnson_2021} was employed assuming a Chabrier IMF \citep{Chabrier_2003} and a two-component dust model \citep{Charlot_2000, Conroy_2009}, while using fixed spec-$z$, as well as including nebular emission. Likewise, $\rm SFR_{10}$ was derived using the \cite{Reddy_2018} calibration, and $\rm 12+log(O/H)$ was determined as in \cite{Laseter_2023}. Regardless of the JADES specifics, we are combining various assumptions involved in deriving $\rm M_{\star}$, $\rm SFR_{10}$, and $\rm 12+log(O/H)$, similar to our $z \sim 0$ combined sample. This combination is a clear limitation moving forward considering the effects different SED assumptions have in the high-$z$ Universe compared to the local Universe \citep[e.g.,][]{Harvey_2025}, but we nonetheless attempt a preliminary comparison before a self-consistent investigation (Laseter et al. in prep.). There is a general agreement in $\rm M_{\star},~SFR_{10}$ and $\rm 12+log(O/H)$ between the JADES, literature, and DJA samples nonetheless. In total, we investigate $\sim 320$ $z \gtrsim 2$ [OIII]$\lambda 4363$ emitters identified with JWST.   

We present in Figure \ref{high-z plot} $\rm \Delta FMR$ \citepalias{Curti_2020} against $\rm 12+log(O/H)$, $\rm M_{\star}$, $\rm SFR_{10}$, $\rm sSFR_{10}$, $f_{\rm gas}$, and $y_{\rm eff}$, along with $y_{\rm eff}$ against $\rm SFR_{10}$. Firstly, the $\rm \Delta FMR$ histogram in panel A demonstrates our high-$z$ sample disproportionally deviates more from the FMR ($\rm \langle \Delta FMR \rangle = -0.57; \langle 12+log(O/H) \rangle = 7.66$) than our low-$z$ sample ($\rm \langle \Delta FMR \rangle = -0.22; \langle 12+log(O/H) \rangle = 7.96$), though sample bias is certainly present. Even so, in panel B we demonstrate the $\rm \Delta FMR - O/H$ relation, from which it is evident that $\rm \Delta FMR \propto log(O/H)$ for both high and low-$z$ systems. In this context, although $\rm \Delta FMR$ still does not scale with $\rm M_{\star}$ (panel C), it is interesting in panels D \& E that high-$z$ systems appear to have stronger $\rm \Delta FMR-SFR_{10}$ and $\rm \Delta FMR-sSFR_{10}$ relationships, in that high-$z$ systems deviate more from the FMR at fixed $\rm sSFR_{10}$. We therefore perform a Welch's t-test for $\rm \Delta FMR$ between the low and high-$z$ samples, simultaneously including observational error using a Monte-Carlo technique. We evaluate the $\rm t-score$ and $\rm p-score$  $\rm 10,000$ times using values drawn randomly from normal distributions for the observational error comprising $\rm \Delta FMR_{err}$. We find a $\rm t-value$ of $6.05$ with a $\rm p-value$ of $\rm 1.45\times10^{-8}$, indicating a statistically significant metallicity deviation. 

This result is in line with the findings of \cite{Curti_2023b}, who presented the $\rm sSFR_{10}$ residuals of $\rm \Delta FMR-\it z$, finding a general trend of higher $\rm sSFR_{10}$ with decreasing $\rm \Delta FMR$. A caveat here is the use of strong-line calibrated metallicities in \cite{Curti_2023b}, but \cite{Heintz_2023} and \cite{Morishita_2024} demonstrated low-$\rm M_{\star}$, pure [OIII]$\lambda 4363$ MZRs are offset to lower metallicities at fixed $\rm M_{\star}$ relative to $z \approx 0$ blueberry and green pea galaxies, which are metal-poor and have comparable excitation properties to the high-$z$ sample \citep{Cameron_2023}. However, if $\rm SFR_{10}$ and $\rm M_{\star}$ are systematically overestimated/underestimated because of poor SED assumptions then this difference could be resolved. However, the collected $\rm M_{\star}$ values are not under the same set of SED assumptions, so a ubiquitous underestimate is unexpected, supported by the high-$z$ $\rm M_{\star}$ distribution in panel C. Likewise, we initially re-derived all $\rm SFR_{10}$ (depending on the availability of H$\alpha$ and H$\beta$) using the conversion of \cite{Reddy_2018} appropriate for low-metallicity galaxies. This relation has a lower $\rm L_{H\alpha}-SFR_{10}$ conversion factor than what is typically used \citep[e.g.,][]{Hao_2011}, and thus a systematic overestimate is unlikely. Sample bias to lower metallicity systems could also be present due to the [OIII]$\lambda 4363$ requirement, but \cite{Laseter_2023}, \cite{Sanders_2023}, and \cite{Scholte_2025} demonstrated a ``flattening" of oxygen strong-line calibrations with $\rm T_{e}$-derived metallicities, suggesting a more uniform $\rm 12+log(O/H)$ distribution at comparable excitation/ionization conditions. It is unlikely then that our metallicity methods are systematically driving high-$z$ systems to lower metallicities. Therefore, considering the $\rm 12+log(O/H)$ and $\rm SFR_{10}$ trends with $\rm \Delta FMR$, high-$z$ systems appear to be statistically more metal-poor than low-$z$ systems at similar $\rm M_{\star}$.

A more robust sample of high-$z$ galaxies within a self-consistent analysis is warranted prior to any strong inferences of this metallicity offset (Laseter et al. in prep.). If real, however, \cite{Curti_2023b} noted a few interpretations all dealing with non-equilibrium conditions and timescales relative to the local Universe: 1) more prominent metal-dilution through accretion of pristine gas, 2) increased levels of stochastic star formation and star formation efficiencies, and 3) enriched outflow efficiency enhancement due to the increase star formation stochasticity and efficiency. We therefore return to comparing metallicity variations relative to the gas-reservoir, i.e., $y_{\rm eff}$, for our high-$z$ systems. To remain consistent, we derive $f_{\rm gas}$ applying the same $f_{\rm gas}-\rm sSFR_{10}$ relation from \cite{Boselli_2014}. We see in panel F these high-$z$ systems are gas-rich, which is expected from observations \citep[e.g.,][]{Saintonge_2016, Scoville_2017, Tacconi_2018, Tacconi_2020} and simulations \citep[e.g.,][]{Popping_2014}, though some systematics could be present from employing a $z\approx 0$ empirical relationship to the high-$z$ Universe. However, it is evident in panel G that high-$z$ systems are associated with lower $y_{\rm eff}$ at fixed $\rm SFR_{10}$. An overestimate of $f_{\rm gas}$ acts to \textit{increase} $y_{\rm eff}$ due to the $1/f_{\rm gas}$ dependence in the denominator of Equation \ref{Effective Yield Equation}, so biasing $y_{\rm eff}$ low requires the \cite{Boselli_2014} relation to substantially \textit{overestimate} $f_{\rm gas}$ in the high-$z$ Universe, which we find unlikely. We therefore expect $y_{\rm eff}$ to vary in a similar fashion as described in \cite{Dalcanton_2007}.

We present in panel H (Figure \ref{high-z plot}) the $y_{\rm eff}-\rm \Delta FMR$ distribution for our high-$z$ systems. In this parameter space, an increase in $y_{\rm eff}$ at fixed $\rm \Delta FMR$ is due to an increase in $f_{\rm gas}$ considering the $\rm \Delta FMR - O/H$ relation (panel B). It is no surprise then the gas-rich, high-$z$ systems are at higher $y_{\rm eff}$ at fixed $\rm \Delta FMR$. Here, we simply see a higher density of high-$z$ systems occupying more negative $\rm \Delta FMR$ values (more metal-poor) and more positive $y_{\rm eff}$ (more gas-rich), which is ultimately panel F re-parameterized. Crucially, we find the slope of $y_{\rm eff}-\rm \Delta FMR$ to align between the low and high-$z$ samples, spanning $y_{\rm eff}$ values across $\rm \sim 3~dex$. As before, gas inflows are ineffective at lowering $y_{\rm eff}$ in gas-rich systems, and so the wide range of our high-$z$ $y_{\rm eff}$ measurements suggests metal dilution from pristine gas is not a dominant driver behind $\rm 12+log(O/H)$ reductions and subsequent FMR deviations (Panel C, Figure \ref{Effective Yields}). We are not implying that infall has not occurred intermittently, but because $y_{\rm eff}$ captures metallicity variations relative to the gas-reservoir, an already rich reservoir with a given metal content is not substantially diluted further. In contrast, enriched outflows are extremely effective in reducing $y_{\rm eff}$ without dramatically changing $f_{\rm gas}$. For example, \cite{Dalcanton_2007} states, ``An enriched outflow that removes less than a fifth of a galaxy’s gas can drop the effective yield by more than a factor of ten, provided that more than half of the galaxy’s baryonic mass is gaseous." 

Nonetheless, our high-$z$ sample is biased to SFGs as nebular emission is required for $\rm T_{e}$-derived metallicities, and thus the high-$z$ inter-burst galaxies (``mini-quenched", ``lulling", ``smoldering", etc.) identified with JWST \citep[e.g.,][]{Looser_2023, Looser_2024, Sun_2023b, Sun_2023} are excluded. The gas content of these systems are unknown and metallicities cannot be directly determined, so we cannot comment on the dominant mechanisms in $y_{\rm eff}$ changes in the context of $\rm \Delta FMR$ as these systems potentially could be gas-poor enough for sudden and massive infall to matter. However, in the gas-rich regime our high-$z$ galaxies occupy, star formation and enriched outflows are the only viable means for the diverse $y_{\rm eff}$ distribution we find. Unenriched infall and outflows are nonetheless present, but they alone cannot describe the metallicity variations relative to a rich but unbalanced gas reservoir in the high-$z$ Universe. 

\subsubsection{\textbf{Supplemental Support (High-$z$)}}

\cite{McClymont_2025}, through their investigation of nitrogen rich galaxies in \texttt{Thesan-zoom} simulations \citep{Kannan_2025}, found low- and high-$z$ systems with $\rm log(M_{\star}/M_{\odot}) \lesssim 9$ deviate from high-$\rm M_{\star}$ systems in $\rm \Delta SFMS - \Delta MZR$ space (Section \ref{Section Non-Parametric}) and that low-$\rm M_{\star}$, high-$z$ systems have a steeper $\rm \Delta SFMS - \Delta MZR$ relation than low-$\rm M_{\star}$, low-$z$ systems, suggesting intrinsically more metal-poor systems at high-$z$. \cite{McClymont_2025} suggests a dominant gas-dilution picture, where FMR deviations are driven by pristine gas inflow diluting the metal content post ISM gas consumption/ejection. We cannot comment on the gas-fractions of inter-burst galaxies, but in Figure \ref{Gas-Fractions} we observe low-$\rm M_{\star}$ systems at $z \sim 0$ remaining gas-rich at lower $\rm SFR_{10}$. However, if high-$z$ galaxies are disproportionally more metal poor due to increased feedback energy then gas-removal may be more prominent. In the extreme case where an outflow completely removes the ISM (e.g., $f_{\rm gas} \approx 0.1$), pristine infall will decrease $y_{\rm eff}$ to a minimum value of around $\rm \sim 1/10th$ of $y_{\rm true}$ \citep{Dalcanton_2007}. The difference between $y_{\rm true}$ and $y_{\rm eff}$ due to pristine infall then diminishes with increasing $f_{\rm gas}$, e.g., a doubling of the gas-mass through infall in a system with $f_{\rm gas} \approx 0.8$ will only decrease $y_{\rm true}$ by $\approx 10\%$ (see Panel C, Figure \ref{Effective Yields}). As such, for our high-$z$ gas-rich systems we simply cannot ascribe the large metallicity and $y_{\rm eff}$ variations to be driven predominately through infall. 

Further theoretical works have suggested canonical FMR departures in the high-$z$ Universe. For example, \cite{Garcia_2024} investigated the FMR in \texttt{Illustris} \citep{Vogelsberger_2014}, \texttt{Illustris-TNG} \citep{Springel_2018}, and \texttt{Eagle} \citep{Schaye_2015} out to $z \sim 8$, introducing the concept of a ``weak" and ``strong"  FMR corresponding to redshift evolution or lack thereof. \cite{Garcia_2024} finds a ``weak" FMR for all simulations, with $\rm \Delta MZR-SFR_{10}$ dependence flattening with increasing $z$, though to what degree depends on the simulation. Interestingly, these results imply the FMR holds to some degree at higher $z$. However, \cite{Garcia_2024} investigated systems with $\rm 8 \leq log(M_{\star}/M_{\odot}) \leq 12$, meaning the entire $\rm M_{\star}$ regime was analyzed together, which could artificially introduce an FMR if the FMR is more established at higher $\rm M_{\star}$ in the early Universe.

\cite{Marszewski_2025} also finds a ``weak" FMR in the high-$z$ suite of FIRE-2 \citep{Hopkins_2018} in four mass bins spanning $\rm 8 \leq log(M_{\star}/M_{\odot}) \leq 10$, in that their lowest $\rm SFR_{10}$ quintiles have higher average metallicities. \cite{Marszewski_2025} demonstrates this dependence disappears when using a longer timescale SFR indicator (e.g., $\rm SFR_{UV}$), and thus suggests the FMR is being driven by pristine infall. However, in the \cite{Marszewski_2025} results, only the lowest $\rm SFR_{10}$ quintile in the mass-complete sample yields any significant deviation; all other $\rm SFR_{10}$ quintiles overlap in their observable sample. Additionally, it is clear in their star formation/metal/gas histories that there is a concurrency between starbursting episodes with metal and gas ejection. Unless there is a consistent dominance of inflows with the same level of enrichment of the recently ejected material, which would have little effect on $y_{\rm eff}$ and be in contention with early galaxy formation models \citep{Dave_2011}, the coeval increases and decreases of the metal content with star formation implies metallicity variations are mainly driven by starbursts and subsequent outflows, not equilibrium nor inflows, resulting in their $\rm SFR_{10}$ quintiles overlapping (i.e., diverse star formation, outflow, and gas balancing conditions). 

Overall, the theoretical ``weak" FMRs discussed generally align with the notion the canonical FMR is not fully applicable in the low-$\rm M_{\star}$ regime in the high-$z$ Universe, but our interpretations of the dominant processes leading to the metallicity variations differ.  

\subsubsection{\textbf{Future Work}}

Within our findings, the metal content of low-$\rm M_{\star}$ systems are not governed by a quasi-steady state gas reservoir and metallicity variations are primarily associated with recent star formation and subsequent outflows. However, it is unclear as to why the high- and low-$z$ Universe differ in the same $\rm M_{\star}$ regime if star formation and enriched outflows dominate. 

The ambient gas metallicity is widely predicted to decrease with increasing $z$ (observationally: \citealt{Fumagalli_2011}, \citealt{Goerdt_2012}; analytically: \citealt{Hernquist_2003}; numerically: \citealt{Wise_2012}), so a possibility is simply the intrinsic metal content of the gas that star formation enriches and evacuates is lower. The analytical findings from \cite{Hernquist_2003} predict a $\rm \sim 0.8~dex$ decrease in the metallicity of ambient gas between $z = 0$ and $z = 6$, which is comparable to the average metallicity offset between our low- and high-$z$ samples ($\rm \langle O/H \rangle_{low-z} - \langle O/H \rangle_{high-z} = 0.3~dex$), especially when considering the analytical models assume a fixed outflow velocity, mass-loss rate, and stellar yield.  

At the same time, the high-$z$ Universe has been shown to be distinct with higher ISM gas temperatures near the limits of standard heating mechanisms \citep[e.g.,][]{Katz_2023, Laseter_2023}, peculiar chemical abundance patterns that are not typically observed in the gas phase \citep[e.g.,][]{Bunker_2023, Cameron_2023b, Isobe_2023, Senchyna_2024, Topping_2024}, and extreme nebular emission \citep[e.g.,][]{Cameron_2024, Katz_2024}, all of which may require exotic stellar populations and more top-heavy IMFs, in addition to increases in the star formation efficiency. There are large theoretical and observational uncertainties in the high-$z$ IMF and supermassive stars though, so it is difficult to gauge the effects these physical processes have on outflows that would lead to additional metallicity decreases. An interesting but contrasting proposal to our picture is the `feedback free starbursts'. For example, \cite{Li_2024} demonstrated that in `feedback free starbursts' the ISM gas content consists of high outflow velocities and \textit{low} gas-fractions ($f_{\rm gas} \approx 0.1$), meaning that pristine infall can substantially affect $y_{\rm eff}$ relative to enriched outflows \citep{Dalcanton_2007}. However, the `feedback free starbursts' picture largely addresses the overabundance of bright galaxies above $z \gtrsim 9$ \citep[e.g.,][]{Finkelstein_2023}, yet our high-$z$ systems consist of more typical SFGs for $ 3 \lesssim z \lesssim 9$, and thus generally below the criteria for `feedback free starbursts'. Additional analysis and discussion is necessitated, but we leave further high-$z$ FMR discussion for forthcoming work in Laseter et al. (in prep.) using a novel high-$z$ simulation \texttt{MEGATRON} \citep{Katz_2024_Mtron, Katz_2025_Mtron}. A $z \sim 0$ analysis was warranted, however, before a \textit{flux-to-flux} comparison of JWST observations with \texttt{MEGATRON} was performed.

\section{\textbf{Summary}} \label{Summary}

We have investigated the FMR in the low-$\rm M_{\star}$ regime ($\rm log(M_{\star}/M_{\odot}) \lesssim 9.0$) for both the low- and high-$z$ Universe; we summarize our findings and methods as follows:

\textit{We find no evidence for the existence of the FMR below $ log(M_{\star}/M_{\odot}) \it \lesssim 9$ through minimization, parametric, and non-parametric techniques.} Regarding minimization, we attempt to minimize the low-$\rm M_{\star}$ MZR with $\rm SFR_{10}$, as is common for standard FMR literature comparisons. We fail to minimize the MZR due to diminished $\rm SFR_{10}$ residual dependencies, i.e., greater $\rm 12+log(O/H)$ dispersion at fixed $\rm SFR_{10}$, ultimately resulting in comparable dispersion to the initial MZR. Parametric-wise, we attempt to fit the form proposed by \cite{Curti_2020}, finding $\rm \langle 12+log(O/H) \rangle$ is better matched but with a metallicity dispersion comparable to the remaining scatter after minimization and to the MZR. Our parameterized fit yields a weak but direct O/H-$\rm sSFR_{10}$ relation that increases with decreasing $\rm M_{\star}$, completely counter to the canonical FMR. Consequently, the standard parameters describing the high-$\rm M_{\star}$ MZR/FMR fail to accurately describe the low-$\rm M_{\star}$ FMR. In our non-parametric investigation, we find no correlation between the location of a low-$\rm M_{\star}$ SFG on the SFMS to its location on the MZR, unlike the inverse correlation identified in high-$\rm M_{\star}$ samples. We find the offset from the FMR is the strongest residual in this parameter space, suggesting the \textit{relative level} of chemical enrichment is more closely related to the FMR offset, thus diminishing the importance of the \textit{relative position} above the SFMS and increasing the respective metallicity dispersion. These results suggest a scenario where star formation rapidly returns metals to the ISM, followed by feedback expelling metals from the ISM prior to subsequent gas-balancing, thus leading to a departure from the canonical FMR in the low-$\rm M_{\star}$ regime.

\textit{We find no correlation between $f_{\rm gas}$, $M_{\star}$, and $\it 12+log(O/H)$ in low-$M_{\star}$ systems, further suggesting gas-processes are deviating from the equilibrium conditions that are shaping the FMR at higher $\rm M_{\star}$.} We find elevated $f_{\rm gas}$ ($\langle f_{\rm gas} \rangle = 0.95$) regardless of metallicity and $\rm SFR_{10}$ (these systems remain gas rich throughout a star formation episode). We also find $f_{\rm gas}$ does not correlate with the offset from the FMR yet does align with the expected trends from the SFMS. These results suggest metallicity variations are happening on shorter timescales than the balancing of gas-rich reservoirs. An example where this condition is encountered is in low-$\rm M_{\star}$ starbursts, as potential disruptions to a steady-state gas-reservoir due to processes such as unbounding the gas, gas-heating, or more efficient star formation are present. Consequently, metallicity of low-$\rm M_{\star}$ systems cannot be ubiquitously described by a gas-regulation model.

\textit{We find that beginning below $\it log(M_{\star}/M_{\odot}) \lesssim 9$, there is a general departure from a steady-state gas reservoir shaping a $\it M_{\star}-SFR_{10}-O/H$ and $\it M_{\star}- f_{\rm gas}- O/H$ relation, in which metallicity variations are primarily driven by star formation and enriched outflows.} We investigate the metal content \textit{relative} to $f_{\rm gas}$ (effective yields; $y_{\rm eff}$), finding $y_{\rm eff}$ directly correlates with FMR offsets due to strong metallicity variations in gas-rich systems. \cite{Dalcanton_2007} demonstrated star formation and enriched outflows, compared to pristine infall, are the only viable means of increasing and decreasing $y_{\rm eff}$, respectively, in gas-rich systems when the reservoir is unbalanced.

\textit{We compare $z \gtrsim 2$ [OIII]$\lambda 4363$ emitters thus far identified with the JWST, finding high-$z$ systems to be disproportionally more metal-poor and offset from the FMR ($\it  \langle12+log(O/H) \rangle = 7.66;  \langle \it \Delta FMR \rangle = -0.57$) than low-$z$ counterparts ($\it \langle 12+log(O/H) \rangle = 7.96; \langle \it \Delta FMR \rangle = -0.22$).} We find the range of $y_{\rm eff}$ values comparable between the low- and high-$z$ Universe, suggesting star formation and enriched outflows still dominate the metallicity variations in high-$z$ gas-rich systems with no steady-state gas-reservoir. However, we cannot conclude if these physical processes lead to additional metallicity decreases as the intrinsic metal content of the gas that star formation enriches and evacuates is predicted to be lower to a comparable degree of our low- and high-$z$ offset. We leave further high-$z$ FMR analysis for forthcoming work in Laseter et al. (in prep.) using a novel high-$z$ simulation \texttt{MEGATRON} \citep{Katz_2024_Mtron, Katz_2025_Mtron}, capable of a direct \textit{flux-to-flux} comparison with JWST observations.

\vspace{-0.8em}
\section{\textbf{Acknowledgments}}
Special thanks to Nicholas Choustikov ({\textcolor{orcidlogocol}{orcid.org/0000-0002-7973-5442}}) for detailed discussions and comments. This material is based upon work supported by the National Science Foundation Graduate Research Fellowship under Grant No. 2137424 and the Fluno Fellowship. MVM is supported by the National Science Foundation via AAG grant 2205519, the Wisconsin Alumni Research Foundation via grant MSN251397, and NASA via STScI grant JWST-GO-4426. AJB \& AJC acknowledge funding from the ``FirstGalaxies" Advanced Grant from the European Research Council (ERC) under the European Union’s Horizon 2020 research and innovation programme (Grant agreement No. 789056). CS acknowledges support from the Science and Technology Facilities Council (STFC), by the ERC through Advanced Grant 695671 “QUENCH”, by the UKRI Frontier Research grant RISEandFALL. Some of the data products presented herein were retrieved from the Dawn JWST Archive (DJA). DJA is an initiative of the Cosmic Dawn Center (DAWN), which is funded by the Danish National Research Foundation under grant DNRF140.
\bibliography{bibliography.bib}
\end{document}